\documentclass[twocolumn]{aastex631}

\newcommand{\lsun}{\mbox{L}_\odot}
\newcommand{\rsun}{\mbox{R}_\odot}
\newcommand{\msun}{\mbox{M}_\odot}
\newcommand{\mearth}{\mbox{M}_\oplus}

\newcommand{\brg}{\mbox{Br}\gamma}
\newcommand{\lacc}{L_{\rm acc}}
\newcommand{\macc}{\dot{M}_{\rm acc}}
\newcommand{\lstar}{L_\star}
\newcommand{\mstar}{M_\star}
\newcommand{\rstar}{R_\star}
\newcommand{\teff}{T_{\rm eff}}
\newcommand{\lbol}{L_{\rm bol}}
\newcommand{\mdisk}{M_{\rm disk}}
\newcommand{\mdust}{M_{\rm dust}}

\usepackage{multirow}
\usepackage{rotating}
\usepackage[graphicx]{realboxes}
\usepackage[maxfloats=256]{morefloats}
\usepackage{tablefootnote}
\usepackage{hyperref}

\received{}
\revised{}
\accepted{\today}

\shorttitle{Accretion on Class I}
\shortauthors{Fiorellino et al.}
\graphicspath{{./}{figures/}}

\begin{document}

\title{The Mass Accretion Rate and Stellar Properties in Class I Protostars}

\author[0000-0002-5261-6216]{Eleonora Fiorellino}
  \affiliation{Konkoly Observatory, Research Centre for Astronomy and Earth Sciences, E\"otv\"os Lor\'and Research Network (ELKH), Konkoly-Thege Mikl\'os \'ut 15-17, 1121 Budapest, Hungary}
  \affiliation{CSFK, MTA Centre of Excellence, Budapest, Konkoly Thege Mikl\'os \'ut 15-17., H-1121, Hungary}
\affiliation{INAF-Osservatorio Astronomico di Capodimonte, via Moiariello 16, 80131 Napoli, Italy}
\author[0000-0002-9470-2358]{{\L}ukasz Tychoniec}
  \affiliation{European Southern Observatory, Karl-Schwarzschild-Strasse 2, 85748 Garching bei München, Germany}

\author[0000-0002-4283-2185]{Fernando Cruz-S\'aenz de Miera }
\affiliation{Konkoly Observatory, Research Centre for Astronomy and Earth Sciences, E\"otv\"os Lor\'and Research Network (ELKH), Konkoly-Thege Mikl\'os \'ut 15-17, 1121 Budapest, Hungary}
  \affiliation{CSFK, MTA Centre of Excellence, Budapest, Konkoly Thege Mikl\'os \'ut 15-17., H-1121, Hungary}
\author[0000-0002-0666-3847]{Simone Antoniucci}
\affiliation{INAF-Osservatorio Astronomico di Roma, via di Frascati 33, 00078, Monte Porzio Catone, Italy}

\author[0000-0001-7157-6275]{\'Agnes K\'osp\'al}
  \affiliation{Konkoly Observatory, Research Centre for Astronomy and Earth Sciences, E\"otv\"os Lor\'and Research Network (ELKH), Konkoly-Thege Mikl\'os \'ut 15-17, 1121 Budapest, Hungary}
  \affiliation{CSFK, MTA Centre of Excellence, Budapest, Konkoly Thege Mikl\'os \'ut 15-17., H-1121, Hungary}
\affiliation{Max Planck Institute for Astronomy, K\"onigstuhl 17, 69117 Heidelberg, Germany}
\affiliation{ELTE E\"otv\"os Lor\'and University, Institute of Physics, P\'azm\'any P\'eter s\'et\'any 1/A, 1117 Budapest, Hungary}

\author[0000-0003-3562-262X]{Carlo F. Manara}
  \affiliation{European Southern Observatory, Karl-Schwarzschild-Strasse 2, 85748 Garching bei München, Germany}

\author[0000-0002-9190-0113]{Brunella Nisini}
\affiliation{INAF-Osservatorio Astronomico di Roma, via di Frascati 33, 00078, Monte Porzio Catone, Italy}

\author[0000-0003-4853-5736]{Giovanni Rosotti}
\affiliation{Leiden Observatory, Leiden University, PO Box 9513, NL-2300 RA Leiden, the Netherlands}
\affiliation{School of Physics and Astronomy, University of Leicester, University Road, Leicester LE1 7RH, UK}
\affiliation{Dipartimento di Fisica 'Aldo Pontremoli', Università degli Studi di Milano, via G. Celoria 16, I-20133 Milano, Italy}

\begin{abstract}

Stars collect most of their mass during the protostellar stage, yet the accretion luminosity and stellar parameters, which are needed to compute the mass accretion rate, are poorly constrained for the youngest sources.
The aim of this work is to fill this gap, computing the stellar properties and the accretion rates for a large sample of Class\,I protostars located in nearby ($< 500$\,pc) star-forming regions and analysing their interplay.
We used a self-consistent method to provide accretion and stellar parameters using SED modeling and veiling information from near-IR observations, when possible. 
We calculated accretion and stellar properties for the first time for 50 young stars. We focused our analysis on the 39 confirmed protostars, finding that their mass accretion rate varies between $\sim 10^{-8}$ and $\sim 10^{-4}$\,$\msun/$yr in a stellar mass range between $\sim 0.1$ and 3\,$\msun$. 
We find systematically larger mass accretion rates for our Class\,I sample than in Class\,II objects. 
Although the mass accretion rate we found is high, it still suggests that either stars collect most of its mass before Class\,I stage, or eruptive accretion is needed during the overall protostellar phase.
Indeed, our results suggest that for a large number of protostars the disk can be unstable, which can result in accretion bursts and disk fragmentation in the past or in the future.

\end{abstract}


\section{Introduction}
\label{sect:intro}

Young stars acquire mass by accreting material from the infalling envelope and the circumstellar disk. 
In particular, according to the magnetospheric accretion scenario \citep{har16}, the accretion flow proceeds from the disk to the forming star along the magnetic field lines.
The accretion rate is supposed to be very large during the protostellar phase (Class\,0 and I), where the accretion luminosity ($\lacc$) is larger than the stellar luminosity ($\lstar$), then it decreases with time in the pre-main sequence (PMS) phase (Class\,II), until accretion basically stops (Class\,III).

In this context, the mass accretion rate ($\macc$) is a fundamental parameter in the star formation process, because it links the properties of the forming star with the evolution of the protoplanetary disk. 
This parameter is well constrained in Classical T\,Tauri (or Class\,II) young stellar objects (YSOs), for which the $\macc$ can be directly measured
from the UV-excess over the stellar photosphere caused by the
accretion shock \citep[e.g.][]{gullbring98, calvet.gullbring98, her08, rig12, ing13, fai15, rugel2018, Schneider2020}.

In the last years, many surveys of the $\macc$ in CTTS have been provided 
\citep[see][for a review]{manara2022pp7}. 
These have been carried out using the UV-excess and/or the empirical correlations between the accretion luminosity ($\lacc$) and the luminosity of accretion tracers, such as the Balmer, Paschen, and Brackett lines \citep[e.g.][]{muz98, alc17}.
On the contrary, the $\macc$ for the earlier stages is known only for a few sources \citep[][]{muz98, nis05a, ant08,Yen.Koch.ea2017}, despite the fact that most of the accretion onto the forming star occurs during the protostellar phase.
The reason behind this is the difficulty of computing accretion rate in such embedded objects. 
Large extinction and veiling of those young protostars prevent us from studying the UV and optical emission, forcing the analysis to be carried out using infrared (IR) wavelengths. 
At those longer wavelengths the contribution of the disk, photosphere, outflows and jets, and envelope are entangled and complicated to be analyzed separately. 

Recent efforts to characterize stellar properties of the youngest stars \cite[e.g.,][]{laos21,fio21} as well as available archival observations \citep{muz98, whi04, dop05, con10} show a promising way to investigate the protostellar phase. 
We continue this effort, aiming to study the accretion process during such early stage of the star formation, by providing results for 39 protostars, enlarging our accretion survey of Class\,I protostars by a factor of 3. 
In this work we will discuss the question: what can we infer about the protostellar accretion process with the current state-of-the-art?

\section{Sample}
\label{sect:sample}

This work is based on the already existing observations of low-mass protostellar sources within 500\,pc published on the literature. We divide the sample into three sub-samples based on how the stellar parameters and the mass accretion rates have been computed.

The first sub-sample consist of 50 protostars for which we calculated stellar parameters and accretion rates for the first time in this work using archival observations (Sect.\,\ref{sect:sample_new});
another sub-sample of 18 protostars is a collection of YSOs with published accretion rates and stellar parameters already present in the literature and directly comparable with the analysis performed in this work (Sect.\,\ref{sect:sample_direct}); lastly, a sub-sample of 27 protostars is collected from the literature, consisting of sources analyzed with indirect methods whose results are not directly comparable with other samples, but that we add for comparison (Sect.\,\ref{sect:sample_undirect}). 

In order to study stellar properties together with protoplanetary disk masses, for all the sub-samples we searched for archival data to collect millimeter fluxes, from which we calculate the disk dust mass ($M_{\rm dust}$).

\subsection{Mass accretion rate and stellar parameters computed in this work}
\label{sect:sample_new}

The first sample is taken from the NASA Infrared Telescope Facility (IRTF) $K$\,band spectroscopic survey of 110 young stars by \citet[][CG10]{con10}. The sources for this survey have been selected from the all-sky {\it IRAS} catalog, classified as Class I  sources by \citet{lada1991} and collected in \citet{Connelley.Reipurth.ea2008a}.

CG10 provided veiling information for 50 out of 110 objects, based on either continuum fitting or photospheric absorption. For the remaining 60 sources, the veiling was too high and a reliable estimate could not be provided. Since veiling is a fundamental parameter in our analysis we narrow down the sample to 50 sources with the veiling measurement available.

Thus we note that in the analysis we are limited toward sources with low veiling value. The maximum veiling value is about $\sim 8$. Since, by definition, large veiling means larger accretion flow, considering only low-veiled young stars implies a bias toward less accreting Class\,I YSOs.

In Tab.\,\ref{tab:sample} we summarize 50 sources from CG10 with their measured properties. In CG10 the source evolution class have been assigned based on \citet{lada1991} classification. We have updated it, considering sources with $-0.3 < \alpha < 0.3$ as Flat spectrum sources \citep[][]{gre94}, in order to highlight the sources in transition between Class\,I and II evolutionary class, however, in further analysis we consider Flat sources and Class I as protostellar sources and we refer to all of them as Class\,I objects.  

Additionally, several objects have been more recently suggested to be more evolved than indicated by their spectral index, based on different protostellar properties \citep{Furlan.McClure.ea2008, Guieu.Dougados.ea2006, Howard.Sandell.ea2013, Furlan.Fischer.ea2016, Sadavoy.Stephens.ea2019}. In particular, four of them (namely IDs: 08, 12, 14, and 35) are Class\,II young stars (see Appendix\,\ref{app:classII}), therefore we discarded them from our sample. 
Six sources (namely IDs: 04, 19, 32, 37, 41, and 43) are eruptive sources classified as FUors (see Appendix\,\ref{app:fuors}). Standard methods of obtaining stellar and accretion propreties are not reliable for FUOri objects, and therefore we consider them separately.

\begin{longrotatetable}
\begin{deluxetable}{lccrcccccccc}
\centering
\tablecaption{\label{tab:sample} Sources selected from  CG10 sample.}
\tablehead{
\colhead{ID} &	IRAS name &	Simbad name	 & $\alpha$ & Class &	R.A.   Dec.				   &	W$_{eq}^{Br\gamma}$& $r_K				$ &  $m_K$ & $\lbol$ &SpT & $\teff$	\\
\colhead{} &             &                 & &                    &     (J2000)      & $\AA$      &                      &    mag & $\lsun$ &           & K}
\decimalcolnumbers
\startdata
1$^b$ &  03220+3035 & [CG2010]IRAS03220+3035(N)  &  0.02 & Flat  & 03:25:09.43  30:46:21.6 & $-1.71 \pm 0.29$ & $   0.0^{+0.88}_{-0.0}$ &  $10.57 \pm 0.05$ & ... & M2 & 3490  \\ 
2 &  03301+3111 & 2MASSJ03331284+3121241		 &  0.31 & I  & 03:33:12.84  31:21:24.1    & $-2.31 \pm 0.08$  & $  0.48^{+0.84}_{-0.3}$ & $10.58 \pm 0.05$ & $ 4.0$ &   M2 & 3490  \\ 
3 &  03507+3801 & IRAS03507+3801 				 &  0.22 & Flat & 03:54:06.19  38:10:42.5  & $-0.64 \pm 0.57$  & $  0.6^{+2.16}_{-0.36}$ & $9.91 \pm 0.06$ & $ 2.5$ &   G7 & 5290  \\ 
4$^F$ &  04108+2803 & NAMEIRAS04108+2803A 		 & $-0.15$ & II (1) & 04:13:53.39 28:11:23.4 & $-1.72 \pm 0.41$  & $ 0.24^{+1.20}_{-0.24}$ & $10.37 \pm 0.02$ & ... &   M2 & 3490  \\ 
5 &  04113+2758 & [BHS98]MHO1 				     & $-0.13$ & Flat & 04:14:26.27  28:06:03.3  & $-3.22 \pm 0.16$  & $  0.60^{+1.94}_{-0.0}$ & $8.20 \pm 0.05$ & $ 1.1$ &   M2 & 3490  \\ 
6$^{\bf b}$ &  04113+2758 & [BHS98]MHO2 		 & $-0.13$ & Flat & 04:14:26.40  28:05:59.7  & $-1.45 \pm 0.20$  & $ 0.00^{+0.72}_{-0.00}$ & $8.27 \pm 0.05$ & $ 1.1$ &   M2 & 3490  \\ 
7 &  04169+2702 & IRAS04169+2702 		         &  0.53 & I & 04:19:58.45  27:09:57.1     & $-1.50 \pm 0.09$ & $ 0.60^{+1.20}_{-0.24}$ &  $11.26 \pm 0.06$ & $ 0.9$ &   M2 & 3490  \\ 
8$^bd?$ &  04181+2655 & [MDM2001]CFHT-BD-Tau19   &  1.96 & II (2) & 04:21:07.95  27:02:20.4 &$-0.89 \pm 0.06$  & $ 0.12^{+3.84}_{-0.12}$ & $10.54 \pm 0.06$ & ... &   G7 & 5290  \\ 
9 &  04181+2655 & NAMEIRAS04181+2654B 		     & ...   & I   & 04:21:10.39  27:01:37.3   & $-8.41 \pm 0.14$  & $ 0.00^{+1.44}_{-0.00}$ & $10.34 \pm 0.06$ & ... &   G7 & 5290  \\ 
10$^{\bf b}$ &  04189+2650 & V*FSTau 			 & ...   & I & 04:22:02.18  26:57:30.5     & $-1.74 \pm 0.06$  & $ 0.72^{+1.20}_{-0.00}$ & $8.28 \pm 0.06$ & $40.0$ &   M2 & 3490  \\ 
11 &  04189+2650 & 2MASSJ04220069+2657324		 & $-0.04$ & Flat & 04:22:00.70  26:57:32.5  & $-6.48 \pm 0.21$  & $     4.9^{+2.8}_-{2.8}$ & $12.03 \pm 0.07 $ & ... &    ... &    ...  \\ 
12 &  04240+2559 & V*DGTau 					     & $-0.26$ & II  & 04:27:04.70  26:06:16.3   & $-7.78 \pm 0.07$  & $ 0.00^{+1.68}_{-0.00}$ & $6.78 \pm 0.05 $ & $ 3.5$ &   G3 & 5740  \\ 
13 &  04248+2612 & IRAS04248+2612 				 &  0.52 & I & 04:27:57.30  26:19:18.4     & $-2.75 \pm 0.25$  & $ 0.00^{+2.76}_{-0.00}$ & $10.62 \pm  0.05$ & $ 0.3$ &   M3 & 3360  \\ 
14 &  04292+2422 & Haro6-13 					 &  0.01 & II (3) & 04:32:15.41  24:28:59.7 &$-2.62 \pm 0.11$  & $ 0.12^{+2.52}_{-0.00}$ & $7.63 \pm 0.05$ & $ 0.6$ &   G7 & 5290  \\ 
15 &  04295+2251 & IRAS04295+2251 				 &  0.13 & Flat & 04:32:32.05  22:57:26.7  & $-1.10\pm 0.08$  & $  0.54^{+1.08}_{-0.30}$ &  $10.54 \pm 0.07$ & $ 0.3$ &   M2 & 3490  \\ 
16 &  04315+3617 & IRAS04315+3617 				 & $-0.05$ & Flat & 04:34:53.22  36:23:29.2  & $-10.22\pm 0.43$ & $    3.6^{+2.6}_{-2.1}$ &  $9.20 \pm 0.05$ & $ 1.7$ &    ... &    ...  \\ 
17$^{\bf b}$ &  04381+2540 & IRAS04381+2540 	 &  0.64 & I & 04:41:12.68  25:46:35.4     & $-2.30\pm 0.18$  & $ 0.24^{+0.66}_{-0.24}$ &  $11.81 \pm 0.05$ & $ 0.6$ &   K5 & 4140  \\ 
18 &  04530+5126 & V*V347Aur 					 &  0.05 & Flat  & 04:56:57.02  51:30:50.9 & $-2.47\pm 0.10$  & $ 0.36^{+1.32}_{-0.00}$ &  $7.80 \pm 0.06$ & ... &   M2 & 3490  \\ 
19$^F$ &  04591-0856 & IRAS04591-0856 			 &  0.62 & I & 05:01:29.64 -08:52:16.9 & $-1.41\pm 0.21$  & $ 0.18^{+0.06}_{-0.18}$ &  $10.40 \pm 0.07$ & $ 0.9$ &   K7 & 3970  \\ 
20 &  05289-0430 & IRAS05289-0430 				 &  0.38 & I & 05:31:27.09 -04:27:59.4     & $-2.85\pm 0.08$  & $    8.0^{+5.0}_{-3.5}$ &  $9.61 \pm 0.05$ & $ 7.1$ &    ... &    ...  \\ 
\enddata
\tablecomments{\begin{footnotesize}$\teff$ from \citet{pec13} for the SpT. Additional properties: {\bf b}inarity, {\bf F}Uor or {\bf E}Xor object \citep[][]{con10}. bAL = alma binary. Class - evolutionary class of the object, classified according to their IR spectral index \citep{lada1991,gre94}. In cases where the classification has been found to be likely different according to other indicators, a reference is provided: 1 -- \cite{Furlan.McClure.ea2008},  2 -- \cite{Guieu.Dougados.ea2006}, 3 -- \cite{Howard.Sandell.ea2013}, 4 -- \cite{Furlan.Fischer.ea2016},  5 -- \cite{Sadavoy.Stephens.ea2019}. The bolometric luminosity $\lbol$, the spectral type (SpT) and the effective temperature ($T_{\rm eff}$) are only shown for comparison with our results in Tabs.\,\ref{tab:first_res} and \ref{tab:results}.\end{footnotesize}}
\end{deluxetable}
\end{longrotatetable}

\begin{longrotatetable}
\begin{deluxetable}{lccrcccccccc}
\tablenum{1}
\centering
\tablecaption{\label{tab:sample} Sources selected from  CG10 sample.}
\tablehead{
\colhead{ID} &	IRAS name &	Simbad name	 & $\alpha$ & Class &	R.A.   Dec.				   &	W$_{eq}^{Br\gamma}$& $r_K				$ &  $m_K$ & $\lbol$ &SpT & $\teff$	\\
\colhead{} &             &                 & &                    &     (J2000)      & $\AA$      &                      &    mag & $\lsun$ &           & K}
\decimalcolnumbers
\startdata
21 &  05311-0631 & 2MASSJ05333251-0629441		 &  0.23 & Flat & 05:33:32.52 -06:29:44.2  & $-2.79\pm 0.10$  & $ 1.32^{+1.32}_{-0.48}$ &  $10.14 \pm 0.07$ & $ 7.3$ &   M4 & 3160  \\ 
22 &  05357-0650 & Parenago2649 				 &  0.01 & Flat & 05:38:09.31 -06:49:16.6  & $-0.72\pm 0.04$  & $ 5.32^{+0.00}_{-3.78}$ &  $ 7.93 \pm 0.03$ & $10.8$ &   A0 &    ...  \\ 
23$^b$ &  05375-0040 & Haro5-90 				 &  0.62 & I & 05:40:06.79 -00:38:38.1     & $-2.59\pm 0.07$  & $ 0.00^{+1.96}_{-0.00}$ &  $ 8.51 \pm 0.02$ & $ 7.1$ &   G7 & 5290  \\ 
24$^b$ &  05375-0040 & 2MASSJ05400637-0038370	 & ...   & I & 05:40:06.37 -00:38:37.0     & $-1.61\pm 0.08$  & $ 0.42^{+0.70}_{-0.28}$ &  $10.28 \pm 0.03$ & ... &   M2 & 3490  \\ 
25$^b$ &  05375-0040 & 2MASSJ05400579-0038429	 & ...& I & 05:40:05.79 -00:38:43.0     & $-5.77 \pm 0.07$ & $ 0.14^{+1.26}_{-0.14}$ &  $ 9.46 \pm 0.02$ & ... &   G7 & 5290  \\ 
26 &  05379-0758 & 2MASSJ05402054-0756398		 & ...   & I & 05:40:20.55 -07:56:39.9     & $-6.10 \pm 0.10$  & $    5.0^{+2.0}_{-5.0}$ & $ 9.35 \pm 0.03$ & $ 6.4$ &    - &    ...  \\ 
27 &  05379-0758 & [MB91]54 					 &  0.19 & Flat & 05:40:20.31 -07:56:24.9  & $-2.35 \pm 0.19$  & $ 0.12^{+1.02}_{-0.12}$ & $10.86 \pm 0.03$ & ... &   K2 & 4760  \\ 
28$^{bAL}$ &  05384-0808 & 2MASSJ05405059-0805487	 &  1.03 & Flat (4) & 05:40:50.59 -08:05:48.7 &$-1.80 \pm 0.14$  & $ 0.30^{+1.02}_{-0.30}$ & $11.48 \pm 0.04$ & $10.8$ &   M2 & 3490  \\ 
29 &  05384-0808 & 2MASSJ05404991-0806084		 & ...   & Flat (4) & 05:40:49.92 -08:06:08.4 &$-2.73 \pm 0.15$  & $ 0.54^{+1.50}_{-0.06}$ & $11.15 \pm 0.04$ & ... &   M2 & 3490  \\ 
30 &  05405-0117 & IRAS05405-0117 				 &  0.71 & Flat (4) & 05:43:03.06 -01:16:29.2 &$-3.06 \pm 0.20$  & $ 1.56^{+1.08}_{-1.56}$ & $10.25 \pm 0.05$ & $ 4.4$ &   M2 & 3490  \\ 
31 &  05427-0116 & IRAS05427-0116 				 &  0.63 & I & 05:45:17.31 -01:15:27.6     & $-0.26 \pm 0.28$  & $ 0.36^{+1.08}_{-0.24}$ & $10.92 \pm 0.05$ & $ 2.5$ &   M2 & 3490  \\ 
32$^{\bf E}$ &  05513-1024 & V*V1818Ori 		 &  0.18 & Flat & 05:53:42.55 -10:24:00.7  & $-2.02 \pm 0.10$  & $    8.8^{+4.8}_{-3.2}$ & $ 5.96 \pm 0.02$ & ... &   F0 & 7280  \\
33$^{\bf b}$ &  05555-1405 & 2MASSJ05574946-1405278	&  0.62 & I &05:57:49.46 -14:05:27.8   & $-8.14 \pm 0.13$  & $    4.9^{+4.4}_{-4.9}$ & $10.62 \pm 0.05$ & $ 4.8$ &    ... &    ...  \\ 
34 &  05555-1405 & 2MASSJ05574918-1406080		 & ...   & I & 05:57:49.18 -14:06:08.0     & $-1.09 \pm 0.08$  & $ 0.54^{+1.14}_{-0.48}$ & $10.86 \pm 0.05$ & ... &   M2 & 3490  \\ 
35 &  16240-2430 & WL16 						 &  0.24 & II (5) & 16:27:02.34 -24:37:27.2 &$-8.09 \pm 0.08$  & $    0.8^{+0.4}_{-0.8}$ & $ 7.82 \pm 0.06$ & ... &   A0 &    ...  \\ 
36$^b$ &  16288-2450 & [CG2010]IRAS16288-2450(W1)	 &  0.70 & I & 16:31:52.98 -24:56:24.6 & $-2.71 \pm 0.18$  & $ 1.20^{+1.08}_{-1.08}$ & $ 7.85 \pm 0.05$ & ... &   M2 & 3490  \\ 
37$^F$ &  16289-4449 & V346 Nor 				 & $-0.04$ & Flat & 16:32:32.19 -44:55:30.7 & $-2.26 \pm 0.11$  & $    6.3^{+4.0}_{-4.3}$ & $ 7.21 \pm 0.08$ & $ 5.9$ &    ... &    ...  \\ 
38 &  16316-1540 & JCMTSFJ163429.4-154700		 &  0.84 & I & 16:34:29.29 -15:47:01.9     & $-0.64 \pm 0.07$  & $ 0.00^{+1.32}_{-0.00}$ & $ 8.28 \pm 0.05$ & $11.4$ &   K0 & 5030  \\ 
39 &  16442-0930 & 2MASSJ16465826-0935197 		 &  0.22 & Flat & 16:46:58.27 -09:35:19.7  & $-2.99 \pm 0.34$  & $    1.4^{+1.4}_{-1.2}$ & $10.92 \pm 0.03$ & $ 0.7$ &    ... &    ...  \\ 
40 &  18275+0040 & IRAS18275+0040 	 			 & $-0.19$ & Flat & 18:30:06.17  00:42:33.6  & $-1.83 \pm 0.08$  & $  11.2^{+25.0}_{-5.0}$ & $ 7.72 \pm 0.06$ & $ 3.4$ &    ... &    ...  \\ 
\enddata
\tablecomments{\begin{footnotesize}$\teff$ from \citet{pec13} for the SpT. Additional properties: {\bf b}inarity, {\bf F}Uor or {\bf E}Xor object \citep[][]{con10}. bAL = alma binary. Class - evolutionary class of the object, classified according to their IR spectral index \citep{lada1991,gre94}. In cases where the classification has been found to be likely different according to other indicators, a reference is provided: 1 f-- \cite{Furlan.McClure.ea2008},  2 -- \cite{Guieu.Dougados.ea2006}, 3 -- \cite{Howard.Sandell.ea2013}, 4 -- \cite{Furlan.Fischer.ea2016},  5 -- \cite{Sadavoy.Stephens.ea2019}. The bolometric luminosity $\lbol$, the spectral type (SpT) and the effective temperature ($T_{\rm eff}$) are only shown for comparison with our results in Tabs.\,\ref{tab:first_res} and \ref{tab:results}.\end{footnotesize}}
\end{deluxetable}
\end{longrotatetable}

\begin{longrotatetable}
\begin{deluxetable}{lccccccccccc}
\tablenum{1}
\centering
\tablecaption{\label{tab:sample} Sources selected from  CG10 sample.}
\tablehead{
\colhead{ID} &	IRAS name &	Simbad name	 & $\alpha$ & Class &	R.A.   Dec.				   &	W$_{eq}^{Br\gamma}$& $r_K				$ &  $m_K$ & $\lbol$ &SpT & $\teff$	\\
\colhead{} &             &                 & &                    &     (J2000)      & $\AA$      &                      &    mag & $\lsun$ &           & K}
\decimalcolnumbers
\startdata
41$^F$ &  18341-0113 & 2MASSJ18364633-0110294    &  0.91 & I & 18:36:46.33 -01:10:29.5 & $-1.66 \pm 0.55$  & $ 0.12^{+2.16}_{-0.00}$ & $ 9.80 \pm 0.05$ & ... &   M2 & 3490  \\ 
42$^{\bf b}$ &  18577-3701 & V*SCrA 			 & 0.12 & Flat & 19:01:08.61 -36:57:20.1    & $-6.50 \pm 0.07$  & $    4.8^{+0.5}_{-2.9}$ & $6.11 \pm 0.05$ & $ 1.5$ &  ... &    ...  \\ 
43$^F$ &  19266+0932 & Parsamian21 				 & 0.37 & I & 19:29:00.86  09:38:42.9      & $-0.18 \pm 0.28$  & $ 0.00^{+2.28}_{-0.00}$ & $ 9.68 \pm 0.07$ & $ 3.4$ &    3 & 5740  \\ 
44 &  20355+6343 & LDN1100 					     & 0.59 & I & 20:36:22.86  63:53:40.4 &      $-1.65 \pm 0.10$  & $     3.9^{+30}_{-3.9}$ & $10.40 \pm 0.05$ & $ 2.5$ &  ... &   ...  \\ 
45 &  21445+5712 & IRAS21445+5712 				 & 0.54 & I & 21:46:07.12  57:26:31.8 &      $-0.63 \pm 0.12$  & $    5.1^{+2.6}_{-3.3}$ & $10.25 \pm 0.13$ & $18.5$ &  ... &  ...  \\ 
46 &  22266+6845 & IRAS22266+6845 				 & 0.53 & I & 22:28:02.99  69:01:16.7 &      $-1.82 \pm 0.08$  & $ 0.12^{+1.56}_{-0.12}$ & $10.49 \pm 0.05$ & $ 1.8$ &   K2 & 4760  \\ 
47 &  22272+6358 & IRAS22272+6358B				 & 1.76 & I & 22:28:57.60  64:13:37.5 &      $-2.42 \pm 0.09$  & $   0.0^{+4.44}_{-0.0}$ & $ 8.18 \pm 0.05$ & $15.5$ &   F3 & 6660  \\ 
48 &  22324+4024 & EM*LkHA233 					 & 0.08 & Flat & 22:34:41.01  40:40:04.5 &   $-2.19 \pm 0.06$  & $    8.4^{+2.2}_{-1.3}$ & $ 9.46 \pm 0.10$ & $11.6$ &   F3 & 6660  \\ 
49 &  23037+6213 & 2MASSJ23054976+6230011 		 & 1.23 & I & 23:05:49.76  62:30:01.2 &      $-2.90 \pm 0.09$  & $  0.60^{+4.44}_{-0.12}$ & $ 9.04 \pm 0.10$ & $30.2$ &   F3 & 6660  \\
50 &  23591+4748 & 2MASSJ00014325+4805189		 & 0.60 & I & 00:01:43.25  48:05:19.0 &      $-2.81 \pm 0.19$  & $  0.84^{+1.32}_{-0.84}$ & $10.42 \pm 0.05$ & ... &   M2 & 3490
\enddata
\tablecomments{\begin{footnotesize}$\teff$ from \citet{pec13} for the SpT. Additional properties: {\bf b}inarity, {\bf F}Uor or {\bf E}Xor object \citep[][]{con10}. bAL = alma binary. Class - evolutionary class of the object, classified according to their IR spectral index \citep{lada1991,gre94}. In cases where the classification has been found to be likely different according to other indicators, a reference is provided: 1 -- \cite{Furlan.McClure.ea2008},  2 -- \cite{Guieu.Dougados.ea2006}, 3 -- \cite{Howard.Sandell.ea2013}, 4 -- \cite{Furlan.Fischer.ea2016},  5 -- \cite{Sadavoy.Stephens.ea2019}. The bolometric luminosity $\lbol$, the spectral type (SpT) and the effective temperature ($T_{\rm eff}$) are only shown for comparison with our results in Tabs.\,\ref{tab:first_res} and \ref{tab:results}.\end{footnotesize}}
\end{deluxetable}
\end{longrotatetable}

This results in a final sample of 40 Class I protostars, for only 39 of them we were able to estimate the stellar and accretion parameters (Sect.\,\ref{sect:analysis}).

\subsection{Mass accretion rate and stellar parameters from the literature: direct measurements}
\label{sect:sample_direct}

In addition to the sample from CG10, we have compiled archival information on Class I sources from the literature for which the accretion rates and stellar parameters were computed using a similar methodology we adopted in this work (see Sect.\,\ref{sect:acc_par}), based on near-IR (NIR) accretion tracers. This includes 18 Class\,I YSOs from \citet{nis05a, ant08} and \citet{fio21}. 
\citet[][]{nis05a} computed the accretion luminosity by subtracting the stellar luminosity from the bolometric luminosity, and showing that this value is in agreement with empirical relations linking HI emission lines and $\lacc$ by \citet[][]{muz98}. \citet[][]{ant08} used this result to build up the same self-consistent method we use in this work (see Sect.\,\ref{sect:acc_par}) by assuming 1\,Myr as the age of their protostars.
\citet{fio21} updated the same aforementioned self-consistent method by using the most recent empirical relations by \citet[][]{alc17} and assuming the age of these sources between the birthline \citep[][]{Palla.Stahler1990} and 1\,Myr old, considering the {\it Spitzer}-based lifetimes \citep[][]{eno09,dun14}.
Table\,\ref{tab:clI_literature} shows the list of these protostars together with their stellar and accretion properties, which are directly comparable with results of the sample described in the previous section. 

\begin{splitdeluxetable*}{ccccccccccBccccc}
\tablecaption{Class\,I previously known from the literature analyzed with the same or a similar method used for this work.\label{tab:clI_literature}}
\tablewidth{0pt}
\tablehead{
\colhead{ID}  & name &	distance & $\lbol$ & $r_K$ & $A_V$ & $\lacc$ & $\lstar$ & $T_{eff}$ & $\mstar$ & $\rstar$ & age & $\macc$ & $M_{\rm dust}$ & ref\\
\colhead{ }&  & pc & $\lsun$ & & $mag$ & $\lsun$ & $\lsun$ & $K$ & $\msun$ & $\rsun$ & $yr$ & $\msun/yr$ & M$_\Earth$ & }
\decimalcolnumbers
\startdata
1 & IRS2	              & 160          & 12	& $-$  & $-$ & $7.7 \pm 2.5$ & $4.3 \pm 1.5$ & $4900 \pm 200$ & $1.4 \pm 0.3$ &	$2.9 \pm 0.5$ & $0.5 - 1  $ & $3\times 10^{-7}$	& $694\pm2$ &  (a)    \\
2 & IRS5a	 	 	      & 160          & 2     & $-$ & $-$ & $0.4	 	   $ & $1.6 \pm 0.5$ & $4200 \pm 200$ & $0.5 \pm 0.1$ &	$1.0 \pm 0.1$ & $0.3 - 0.5$ & $3\times 10^{-8}$  & $587\pm2$  & (a)    \\
3 & IRS6a	 	 	      & 160          & 0.3	& $-$ & $-$ & $< 0.1      $ & $0.5 \pm 0.2$ & $3580 \pm 100$ & $0.3 \pm 0.1$ &	$0.1 \pm 0.1$ & $0.5 - 1  $ & $< 5\times 10^{-9}$ & -  & (a)   \\
4 & HH100IR	              & 160          & 14	& $-$ & $-$ & $12 \pm   2 $ & $3.1 \pm 0.9$ & $4060 \pm 300$ & $0.4 \pm 0.1$ &	$6.0 \pm 0.5$ & $0.1	  $ & $2\times 10^{-6}$  & $444\pm8$ & (a)\\
5 & IRS3		          & 160          & 0.3	& $-$ & $-$ & $< 0.1 	   $ & $0.3 \pm 0.1$ & $3800 \pm 100$ & $0.5 \pm 0.1$ &	$0.2 \pm 0.1$ & $1 - 5	  $ & $< 9.6 \times 10^{-9}$ & - & (a) \\
6 & HH 26 IRS             &	450          & $4.6 - 9.2$ &  $-$  & 4.1 & 3.2  & 3.7 & K7	& 0.6 & $-$ & $-$ & $8.5 \times 10^{-7}$  & $469\pm18 $ & (b) \\
7 & HH 34 IRS             &	460          & $12.4 - 19.9$  & $-$ & 4.9 	  & 13.3 & 2.9 & M0	& 0.5 & $-$ & $-$ & $41.1 \times 10^{-7}$ & $ 2283\pm68 $ &(b) \\
8 & HH 46 IRS             &	450          & $< 15.0$  & $-$ &  4.9 	  & 1.5  & 6.0 & K5	& 1.2 & $-$ & $-$ & $2.2 \times 10^{-7}$  &$ 1127\pm15 $ & (b) \\
9 & J03283968+3117321$^*$ & $293 \pm 22$ & 0.31 & $3-8$  & $43.0 - 51.0$  & $0.04 - 0.10$  & $0.25-0.31$  & $2818-3038$  & $0.13 - 0.20$  & $1.6 - 2.2$ & BL - 1Myr & $(1.4-8.3) \times 10^{-8}$ & $12.8\pm0.8$ & (c) \\ 
10 & J03285842+3122175$^*$ & $293 \pm 22$ & 1.11 & $3-8$  & $31.0 - 34.5$  & $0.57 - 0.86$  & $0.25-0.54$  & $3020-3218$  & $0.11 - 0.25$  & $1.8 - 2.5$ & BL - 1Myr & $(19-70  ) \times 10^{-8}$ & $11.3\pm0.6$ & (c)\\ 
11 & J03290149+3120208$^*$ & $293 \pm 22$ &18.2  & $3-8$  & $45.0 - 52.0$  & $2.46 - 5.70$  & $12.5-15.7$  & $4942-5248$  & $2.80 - 3.03$  & $4.5 - 4.9$ & BL - 1Myr & $(16-38  ) \times 10^{-8}$ & $7.7\pm2.2$  &  (c)\\ 
12 & SVS 13 (V512 Per)$^{\dagger}$ & $293 \pm 22$&58.8  & $3-8$  & $33.0 - 39.0$  & $16.4 - 33.7$  & $25.1-42.5$  & $5173-5754$  & $3.07 - 3.53$  & $2.1 - 5.7$ & BL - 1Myr & $(19-220 ) \times 10^{-8}$ & $969.7\pm15.5$& (c)\\ 13 & LAL96 213             & $293 \pm 22$ & 7.63 & $3-8$  & $17.0 - 41.0$  & $1.72 - 4.01$  & $3.62-5.91$  & $3802-4683$  & $0.70 - 1.90$  & $3.1 - 4.7$ & BL - 1Myr & $(13-120 ) \times 10^{-8}$ & $318.5\pm0.9$ & (c) \\ 
14 & J03290895+3122562$^*$ & $293 \pm 22$ & 0.23 & $1-3$  & $15.5 - 22.5$  & $0.02 - 0.04$  & $0.19-0.22$  & $2754-3002$  & $0.11 - 0.15$  & $1.4 - 2.1$ & BL - 1Myr & $(0.7-3.3) \times 10^{-8}$ & $-$ & (c) \\ 
15 &  J03290907+3121291$^*$ & $293 \pm 22$ & 0.51 & $3-8$  & $36.5 - 41.0$  & $0.20 - 0.34$  & $0.17-0.31$  & $2754-3075$  & $0.11 - 0.20$  & $1.4 - 2.1$ & BL - 1Myr & $(7-28   ) \times 10^{-8}$ &  $-$&  (c) \\ 
16 & J03291188+3121271$^*$ & $293 \pm 22$ & 0.12 & $1-3$  & $19.0 - 25.5$  & $0.01 - 0.03$  & $0.09-0.11$  & $2754-2928$  & $0.10 - 0.11$  & $1.0 - 2.1$ & BL - 1Myr & $(0.6-2.6) \times 10^{-8}$ & $-$ & (c) \\ 
17 & J03292003+3124076$^*$ & $293 \pm 22$ & 0.59 & $1-3$  & $26.0 - 34.0$  & $0.03 - 0.07$  & $0.52-0.56$  & $3020-3289$  & $0.19 - 0.28$  & $2.1 - 2.5$ & BL - 1Myr & $(0.8-3.6) \times 10^{-8}$ & $3.8 \pm1.6$ & (c) \\ 
18 & J03292044+3118342$^*$ & $293 \pm 22$ & 0.70 & $1-3$  & $12.5 - 20.0$  & $0.06 - 0.15$  & $0.55-0.64$  & $3020-3289$  & $0.19 - 0.28$  & $2.2 - 2.5$ & BL$-$1Myr & $(2.0-8.1) \times 10^{-8}$ & $-$ & (c) \\
\enddata
\tablecomments{($^*$) 2MASS name of the sources (the prefix 2MASS has been removed). (a) -- \citet{nis05a}; (b) -- \citet{ant08}; (c) -- \citet{fio21}. ($^{\dagger}$) binary system. }
\end{splitdeluxetable*}

\subsection{Mass accretion rate and stellar parameters from the literature: indirect measurements}
\label{sect:sample_undirect}

Obtaining direct measurements of stellar properties for the youngest sources is challenging, however, there exists a number of indirect methods to derive stellar and accretion parameters that we describe below. 
In Table\,\ref{tab:indirect_sources}, we collect information on bolometric luminosity, stellar mass and disk dust mass on protostars (9 Class\,0 and 12 Class\,I sources in addition to our 39+18 Class\,I protostars).

The $\mstar$ have been provided by inferring the mass of a central object responsible for the Keplerian rotation of the circumstellar disk \citep[e.g.,][]{Lommen.Joergensen.ea2008, Chou.Takakuwa.ea2014,Lee2010}, except one source WL16 where \cite{Miotello.Testi.ea2014} the $\mstar$ is obtained from $\lbol$ with standard assumptions on the $\macc$ and placing a star on the birthline. We also note that if the estimate depends on the distance (e.g. bolometric luminosity and disk dust mass), we scaled it to the current best estimate for the distance (see Sect.\,\ref{sect:dist}).

It should be noted that other stellar properties such as stellar and accretion luminosity and mass accretion rate can be obtained for some protostars using indirect methods and with assumptions on stellar evolutionary track. 
Compilations of some of those properties for protostars are presented in \cite{Yen.Koch.ea2017} and \cite{Sai.Ohashi.ea2020}, where then-known rotationally-supported disk are listed and their host star properties are summarized.  

We acknowledge that those properties may be less reliable than in direct measurements described in Sections \ref{sect:sample_new} and \ref{sect:sample_direct}. However, this addition not only offers insight into the sources where direct measurements are very challenging (e.g., very extincted sources as Class\,0 and high veiled Class\,I protostars), but also allows to shed some light on the reliability of these indirect methods. Examples of the other indirect methods to derive stellar properties are briefly summarized below.

\cite{dop05} measured $\lstar$ from $H$ and $K$\,band photometry corrected for extinction with caveat that the accretion could contribute to these fluxes. 
\cite{Prato.Lockhart.ea2009} obtained $\lacc$ from Br$\gamma$ flux measurements, $K$\,band photometry, and spectral typing by comparing with standard stars. 
\cite{Lee2010} estimated $\lstar$ using assumption of $\rstar=3\rsun$ and 
$\mstar$ from the Keplerian rotation of the gas in the disk, assuming that the accretion rate onto the star is equal to the infall rate onto the disk. 
In \cite{Miotello.Testi.ea2014}, the effective temperature ($\teff$) is derived from $L_{\rm bol}$ and by placing a star on the birthline from \cite{Palla.Stahler1990}. 
\cite{Sheehan.Eisner2018} obtained stellar luminosity through radiative transfer modeling of the Spectral Energy Distribution (SED). 
From this, a stellar radius can be obtained and, by following assumption of $\macc =$ 10$^{-5}$ $\msun$yr$^{-1}$, a $\mstar$ is provided. 
Alternatively, $\mstar$ can be provided by inferring the mass of a central object responsible for the Keplerian rotation of the circumstellar disk \citep{Lommen.Joergensen.ea2008, Chou.Takakuwa.ea2014,Lee2010}.
The mass accretion rate can be approximated from $\lbol$ where $\rstar$ is assumed to be 3$\rsun$ and $\mstar$ is derived from Keplerian rotation of the disk \citep{Yen.Koch.ea2017}.

\section{Analysis} \label{sect:analysis}
In this section, we describe how we computed the distance, the bolometric luminosity ($\lbol$), and the disk mass ($\mdisk$) for the overall sample. 
We also discuss the computation of the accretion rates and stellar parameters for the sample in Tab.\,\ref{tab:sample}.  
The accretion rates and stellar parameters for sources in Tabs.\,\ref{tab:clI_literature} and \ref{tab:indirect_sources} have already been computed in the literature.

\begin{deluxetable*}{lccccccc}
\tablecaption{Class 0 and I protostars collected from the literature with stellar and accretion properties obtained with alternative methods to those in Table 2 and in this work. \label{tab:indirect_sources}}
\tablewidth{0pt}
\tablehead{
\colhead{ID} & Name & Notes$^{a}$ &	Class & D & $\lbol$ & $\mstar$ & $\mdust$ \\
\colhead{} & &  &  & pc & $\lsun$ & $\msun$  & M$_\Earth$ \\}
\decimalcolnumbers
\startdata
1 & B335            & -                & 0$^1$      & $106\pm15$$^2$     & 0.8$^1$        & $0.05$$^3$ 		        & $14\pm4^4$  			\\
2 & IRAS16253-2429  & -                & 0$^5$      & $144\pm9^6$        & 0.09$^5$       & $0.03\pm0.01^7$       & $9.2\pm1.2^8$ 		\\
3 & VLA1623A        & - 			   & 0$^1$      & $144\pm9^6$        & 4.8$^{1,*}$     & $0.22 \pm 0.02^9$       & $129 \pm 16^{10}$         \\
4 & IRAS15389-3559  & -                & 0$^{11}$   & $155\pm4^{12}$     & 1.5$^{11}$     & $0.007\pm0.004^{13}$    & $5.8\pm0.3^{13}$           \\
5 & Lupus3-MMS      & -                & 0$^{14}$   & $162\pm3^{15}$     & 0.27$^{14,*}$  & $0.3\pm0.1^7$           & $213\pm8^7$ 		    \\
6 & L1455 IRS1      & b$^{16}$ 		   & 0$^{17}$   & $293\pm22^{18}$    & 4.94$^{17}$    & $0.32^{3}$  		    & $175\pm16^{19}$			\\
7 `&IRAS4A2         & -  			   & 0$^1$      & $293\pm22^{18}$    &$<14.1^{1,*,+}$ & $0.08\pm0.02^{20}$      & $1637\pm148^{19}$         \\
8 & L1157           & b$^{21}$ 		   & 0$^1$      & $352\pm19^6$       & 7.9$^{1,*}$    & $0.04^{22}$   		    & $985\pm195^{23}$ 		    \\
9 & HH212           & -                & 0$^{24}$   & $400\pm40^{25}$    & 13.5$^{24}$    & $0.3\pm0.1^{26}$        & $474\pm95^{25}$         \\
10 &L1527           & -                & I$^1$      & $141\pm9^6$        & 2.5$^1$        & $0.19\pm0.04^{27}$      & $221\pm28^{28}$       \\
11 & L1551-IRS5      & b$^{29}$ 		   & I$^1$      & $141\pm9^6$        & 22.9$^{1,*}$   & $0.2 \pm 0.1^{30}$      & $526\pm83^{31}$   		\\
12 & TMC1A           & -                & I$^1$      & $141\pm9^6$        & 2.6$^1$        & $0.56^{32}$  			& $210 \pm 34^{33}$ 		\\
13 & L1489 IRS       & -                & I$^{34}$   & $141\pm9^6$        & 3.5$^{34}$     & $1.64\pm0.12^{35}$ 	    & $52\pm8^{35}$ 		    \\
14 & L1551-NE        & b$^{36}$		   & I$^{37}$   & $141\pm9^6$        & 4.2$^{37}$     & $0.8^{38}$ 		        & $187\pm24^{39}$  			\\
15 & WL12	        & - 			   & I$^{1}$    & $144\pm9^6$        & 1.9$^{1,*}$    & $3.8^{40,*}$ 			& $62\pm8^{41}$  		\\
16 & Elias29	        & -  			   & I$^{1}$    & $144\pm9^6$        & 19.3$^{1,*}$   & $3.2 \pm 0.6^{42}$	    & $15.6\pm2.0^{43}$  		\\
17 & IRS63           & -                & I$^{34}$   & $144\pm9^6$        & 2.0 $^{34,*}$  & 0.8$^{44}$ 		        & $305\pm38^{45}$     	    \\
18 & IRS43           & b$^{46}$ 		   & I$^{14}$   & $144\pm9^6$        & 1.4$^{14,*}$   & 1.9$^{44}$             & $15\pm2^{43}$			    \\
19 & RCrA IRS7B      & b$^{41}$ 		   & I$^{1}$    & $155\pm4^{12}$     & 6.5$^{1,*}$    & $2.3^{47}$  		    & $393\pm17^{41}$           \\
20 & HH111           & b$^{48}$ 		   & I$^{37}$   & $411\pm41^{25}$    & 23$^{37}$     & 1.3$^{49}$ 	 		   & $187 \pm 24$		\\
21 & EC53            & -                & I$^{17}$   & $436\pm9^{50}$     & 5.4$^{17}$    & $0.3\pm0.1^{51}$         & $381\pm16^{51}$		    \\
\enddata
 \tablecomments{$*$: scaled for updated distance; $+$: upper limit, property derived from unresolved observation;
 $a$- b: binary system, the mass of the disk and stellar mass is from the sum of both components.
 {$^1$}\cite{Karska.Kaufman.ea2018}, 
                {$^2$}{\cite{Olofsson.Olofsson2009}},
                $^3$\cite{Yen.Takakuwa.ea2015},
                $^4$\cite{Bjerkeli.Ramsey.ea2019},
                $^5$\cite{dun08},
                $^6$\cite{zuc19},
                $^7$\cite{Yen.Koch.ea2017},
                $^8$\cite{Hsieh.Hirano.ea2019},
                $^{9}$\cite{mur13},
                $^{10}$\cite{Sadavoy.Myers.ea2018},
                $^{11}$\cite{Yang.Green.ea2018},
                $^{12}$\cite{Galli.Bouy.ea2020},
                $^{13}$\cite{Okoda.Oya.ea2018},
                $^{14}$\cite{dun13},
                $^{15}$\cite{Dzib.Loinard.ea2018},
                $^{16}$\cite{tob16},
                $^{17}$\cite{dun15},
                $^{18}$\cite{ort18},
                $^{19}$\cite{Tobin.Looney.ea2018},
                $^{20}$\cite{Choi.Tatematsu.ea2010},
                $^{21}$\cite{Tobin.Cox.ea2022},
                $^{22}$\cite{Kwon.FernandezLopez.ea2015},
                $^{23}$\cite{Chiang.Looney.ea2012},
                $^{24}$\cite{Furlan.Fischer.ea2016},
                $^{25}$\cite{tob20},
                $^{26}$\cite{Codella.Cabrit.ea2014},
                $^{27}$\cite{Tobin.Hartmann.ea2012},
                $^{28}$\cite{Nakatani.Liu.ea2020},
                $^{29}$\cite{Looney.Mundy.ea1997},
                $^{30}$\cite{Chou.Takakuwa.ea2014},
                $^{31}$\cite{CruzSaenzdeMiera.Kospal.ea2019},
                $^{32}$\cite{Aso.Ohashi.ea2015},
                $^{33}$\cite{Harsono.vanderWiel.ea2021},
                $^{34}$\cite{Green.Evans.ea2013},
                $^{35}$\cite{Sai.Ohashi.ea2020},
                $^{36}$\cite{Reipurth.Rodriguez.ea2002},
                $^{37}$\cite{Froebrich2005},
                $^{38}$\cite{Takakuwa.Saito.ea2013},
                $^{39}$\cite{Takakuwa.Saigo.ea2017a},
                $^{40}$\cite[estimated from Lbol, assuming Macc and birhtline][]{Miotello.Testi.ea2014},
                $^{41}$ALMA: 2019.1.01792.S,
                $^{42}$\cite{Lommen.Joergensen.ea2008},
                $^{43}$\cite{Sadavoy.Stephens.ea2019},
                $^{44}$\cite{Brinch.Joergensen2013},
                $^{45}$\cite{SeguraCox.Schmiedeke.ea2020},
                $^{46}$\cite{Girart.Rodriguez.ea2000},
                $^{47}$\cite{Lindberg.Joergensen.ea2014},
                $^{48}$\cite{Reipurth.Yu.ea1999},
                $^{49}$\cite{Lee2010},
                $^{50}$\cite{ort17},
                $^{51}$\cite{Lee.Lee.ea2020},
                }
\end{deluxetable*}

\subsection{The distance} \label{sect:dist}
We verified the distances to our targets with the most recent estimates. 
If available, direct measurement of the parallax from {\it Gaia} EDR3 was used to update the distance \citep{GaiaCollaboration.Brown.ea2021}. 
In many cases the sources in our sample are too deeply embedded to be measured with {\it Gaia} and therefore we adopted the parent cloud distance obtained with {\it Gaia} \citep{Krolikowski.Kraus.ea2021, Akeson.Jensen.ea2019, Yan.Zhang.ea2019, tob20} or other methods \citep{Kospal.Abraham.ea2008, Wouterloot.Brand1989, Hilton.Lahulla1995, Anglada.Rodriguez2002a, ort18}. Additionally for source 2MASSJ05574918-1406080 (ID: 34) a large discrepancy was found between parallax and photogeometric method \citep{bai20} and since the latter has a smaller uncertainty and is more consistent with distance to the other sources nearby, we chose this method instead of the parallax measurement.

\subsection{The bolometric luminosity}
\label{sect:lbol}
The bolometric luminosity is a critical parameter in our analysis.
Calculation of $\lbol$ relies on good sampling of the SED of the protostellar sources. 
We used the SEDBYS v.2.0 python-based package \citep{Davies2021} to extract photometry and build SEDs (Figs. in Appendix\,\ref{app:sed}). 
In addition to the catalogs available in the SEDBYS package, we cross-checked our list of targets with catalogs of dense cores: {\it Herschel} \citep{Marsh.Kirk.ea2016, Pezzuto.Benedettini.ea2021}, {\it Herschel} PACS Point Source Catalog\footnote{https://doi.org/10.5270/esa-rw7rbo7}, {\it Herschel} SPIRE  Point Source Catalog\footnote{https://doi.org/10.5270/esa-6gfkpzh} and photometry catalog of disks in Taurus 
\citep{Andrews.Rosenfeld.ea2013}. 
Using all the available photometry, it was possible to compute $\lbol$ for every source in the sample, and to provide more accurate estimates than what is available in the literature. 
Absence of photometry at any specific wavelengths  means that the source was either not covered by the observations or it was not detected. 
We did not include sub-arcsecond observations of the disks in the SED data as those would usually resolve out the envelope, underestimating the contribution of the envelope at those wavelengths. 

We computed $\lbol$ by integrating the SEDs with a dedicated python procedure, following the method already used in the literature \citep{ant08, fio21}. 
The integration was performed starting from the shortest wavelength, and using linear interpolation in the $\log \lambda - \log (\lambda F_\lambda)$ plane between the available SED points. 
We also applied a final correction at the longest wavelengths ($\lambda > 100\,\mu$m), assuming that the emission decreases as $1/\lambda^2$
after the last available observation. 
Since the available photometry differs source by source, we adopted the following methodologies to provide homogeneous estimate of the bolometric luminosity.
For sources for which photometry up to 2\,mm was observed (36\%), we just integrated the fluxes, as described above.
For sources for which we collected photometry at $\lambda > 100 \, \mu$m but no flux at 2\,mm, we fit a straight line (in log-log) from the peack of the SED to the longest available wavelength to extrapolate the photometry at 2\,mm, and then we proceed with the integration (44\%).
Lastly, several sources had no photometry at wavelengths longer than $100 \, \mu$m in our dataset (20\%). In some cases, this is due to the fact that the source is too faint at such long wavelength, and it is not detected. But in other cases, we are not certain whether the source has a strong emission or not at $\lambda > 100 \, \mu$m since no observations at these wavelegths were performed. 
In these cases we integrated all the available photometry being aware that it is possible that our results are lower limits. In the following, we highlight these sources in the relevant plots.

The derived values are shown in Tab.\,\ref{tab:results} and they range from $\sim 10^{-3}$\,$\lsun$ to $\sim 10^3$\,$\lsun$. 
Uncertainties on $\lbol$ were computed by adopting Monte Carlo simulations.

\startlongtable
\begin{deluxetable*}{lcccccc}
\tablecaption{Computed distance, bolometric luminosity, Br$\gamma$ flux, and dust mass of the sample in Tab.\,\ref{tab:sample}. \label{tab:first_res}}
\tablewidth{0pt}
\tablehead{
\colhead{ID} & Name &	Cloud & Distance & $\lbol$ & $F_{{\rm Br}\gamma}$ &  $\mdust$ \\
\colhead{  } &      &         & pc & $\lsun$ & erg s$^{-1}$ cm$^{-2}$ &  $\mearth$ }
\decimalcolnumbers
\startdata
01 		 & CG2010IRAS032203035N 	& Per-IC348 	& $219.8\pm16.2 $  $^{a}$ & $1.47_{0.21}^{0.26} $ &  $ (4.25 \pm 1.34) \times 10^{-15} $ & $168 \pm 25 $ \\
02 		 & 2MASSJ033312843121241 	& Per-IC348 	& $319.5\pm23.7	$  $^{a}$ & ${5.43_{0.68}^{0.99}}^\dagger $ &  $ (5.69 \pm 7.06) \times 10^{-16} $ & $190 \pm 28$ \\
03 		 & IRAS035073801 			& Per-B1 		& $310.9\pm22.8 $  $^{a}$ & $4.81_{0.64}^{0.75} $ & 	$ (2.92 \pm 3.45) \times 10^{-15} $ & $-$\\
04$^{*}$ & NAMEIRAS041082803A 		& Tau-L1495 	& $170.1\pm24.9 $  $^{a}$ & ${0.30_{0.08}^{0.11}}^\dagger$ &  $ (5.14 \pm 1.91) \times 10^{-15} $ & $-$\\
05 		 & BHS98MHO1 				& Tau-L1495 	& $134.0\pm7.0 $   $^{a}$ & $1.69_{0.16}^{0.14} $ &  $ (7.10 \pm 0.89) \times 10^{-14} $ & $171 \pm 18$ \\
06 		 & BHS98MHO2 				& Tau-L1495 	& $131.0\pm2.9$    $^{a}$ & $1.20_{0.05}^{0.05} $ &  $ (3.00 \pm 0.93) \times 10^{-14} $ & $101 \pm 5$ \\
07 		 & IRAS041692702 			& Tau-L1495 	& $129.5 \pm 12.9 $ $^b$  & $1.09	_{0.21}^{0.19} $ &  $ (1.97 \pm 0.40) \times 10^{-15} $ & $147 \pm  47$ \\
08$^{*}$ & MDM2001CFHTBDTau19 		& Tau-B213 		& $155.9 \pm 15.6 $ $^b$  & $0.63_{0.10}^{0.14} $ &  $ (2.27 \pm 0.63) \times 10^{-15} $ & $7.2 \pm 1.5$\\
09 		 & NAMEIRAS041812654B 		& Tau-B213 		& $155.9\pm 15.6 $ $^b$   & $0.74_{0.13}^{0.17} $ &  $ (2.58 \pm 0.13) \times 10^{-15} $ & $-$\\
10 		 & VFSTau 					& Tau-Aur 		& $133.9\pm2.4$ $^{a}$    & $4.15_{0.16}^{0.10} $ &  $ (3.56 \pm 0.51) \times 10^{-14} $ & $1.8 \pm 0.1$ \\
11 		 & 2MASSJ042200692657324 	& Tau-Aur 		& $133.9\pm2.4$ $^{z}$ 	  & $0.90_{0.08}^{0.05} $ &  $ (4.20 \pm 0.32) \times 10^{-15} $ & $97 \pm  10$ \\
12$^{*}$ & VDGTau 					& Tau-Aur 		& $125.3\pm1.9$ $^{a}$ 	  & $3.89_{0.16}^{0.11} $ &  $ (6.35 \pm 0.25) \times 10^{-13} $ & $325 \pm 11 $ \\
13 		 & IRAS042482612 			& Tau-B213 		& $155.9 \pm 15.6$  $^b$  & $0.69_{0.09}^{0.14} $ &  $ (6.53 \pm 1.19) \times 10^{-15} $ & $-$ \\
14$^{*}$ & Haro613 					& Tau-Aur 		& $128.6\pm1.6$ $^{a}$    & $1.08_{0.03}^{0.03} $ &  $ (9.77 \pm 1.26) \times 10^{-14} $ & $200 \pm 5$ \\
15 		 & IRAS042952251 			& Tau-L1546 	& $160.76 \pm 16.1$ $^b$  & $0.81_{0.16}^{0.21} $ &  $ (2.81 \pm 0.73) \times 10^{-15} $ & $125 \pm 62$ \\
16 		 & IRAS043153617 			& California	& $359.4\pm24.7	$ $^a$ 	  & $3.44_{0.46}^{0.46} $ & 	$ (8.98 \pm 0.60) \times 10^{-14} $ & $-$\\      
17 		 & IRAS043812540 			& Tau-L1527 	& $141.8 \pm 1.4$	$^b$  & $0.60_{0.02}^{0.03} $ &  $ (2.94 \pm 0.55) \times 10^{-15} $ & $25 \pm 5$ \\
18 		 & VV347Aur 				& Tau-Aur 		& $206.6\pm2.3$		$^a$  & $3.49_{0.10}^{0.14} $ &  $ (7.88 \pm 1.03) \times 10^{-14} $ & $-$\\
19$^{*}$ & IRAS045910856 			& ONC A 		& $400	\pm 40 $	$^c$  & $2.95_{1.25}^{1.69} $ & 	$ (4.10 \pm 1.34) \times 10^{-13} $ & $-$\\
20 		 & IRAS052890430 			& ONC A 		& $400	\pm 40 $	$^c$  & $7.78_{1.52}^{1.55} $ & 	$ (1.71 \pm 0.17) \times 10^{-14} $ & $-$\\
21 		 & 2MASSJ053332510629441 	& ONC A 		& $390	\pm 39 $	$^c$  & $14.0_{4.2}^{5.7} $ & 	$ (1.03 \pm 0.12) \times  10^{-14} $ & $-$\\
22 		 & Parenago2649 			& ONC A 		& $398.5\pm2.5 $	 $^a$ & $12.9	_{0.2}^{0.3} $ & 	$ (2.03 \pm 0.57) \times 10^{-15} $ & $ 105 \pm 21$ \\
23 		 & Haro590 					& ONC B 		& $416.4\pm7.8 $     $^a$ & $9.74_{1.41}^{1.49} $ & 	$ (4.29 \pm 0.44) \times 10^{-14} $ & $-$ \\
24 		 & 2MASSJ054006370038370 	& ONC B 		& $408.2\pm12.4	$    $^a$ & ${0.87_{0.06}^{0.05}}^\dagger$ &  $ (5.23 \pm 0.92) \times 10^{-15} $ & $-$ \\
25 		 & 2MASSJ054005790038429 	& ONC B 		& $421.9\pm10.7 $	 $^a$ & ${2.76_{0.17}^{0.17}}^\dagger$ & 	$ (3.99 \pm 0.19) \times 10^{-14} $ & $-$ \\
26 		 & 2MASSJ054020540756398 	& ONC A 		& $427.2\pm72.1 $  $^a$   & $10.4_{4.1}^{3.7} $ & 	$ (4.66 \pm 0.25) \times 10^{-14} $ & $-$ \\
27 		 & MB9154 					& ONC A 		& $502.6\pm183.3 $ $^a$	  & ${0.60_{0.40}^{0.51}}^\dagger$ &  $ (4.47 \pm 0.83) \times 10^{-15} $ & $-$ \\
28 		 & 2MASSJ054050590805487 	& ONC A 		& $440 \pm 44 $	$^c$	  & $1.12	_{0.20}^{0.20} $ &  $ (1.93 \pm 0.40) \times 10^{-15} $ & $72 \pm 15$ \\
29 		 & 2MASSJ054049910806084 	& ONC A 		& $440 \pm 44 $	$^c$	  & $2.63_{0.46}^{0.52} $ & 	$ (3.98 \pm 0.57) \times 10^{-15} $ & $10 \pm 3$ \\
30 		 & IRAS054050117 			& ONC B 		& $420 \pm 42 $	$^c$	  & $3.18_{0.56}^{0.71} $ & 	$ (1.02 \pm 0.15) \times 10^{-14} $ & $ 192 \pm 39$ \\
31 		 & IRAS054270116 			& ONC B 		& $420 \pm 42 $	$^c$	  & ${13.2_{3.1}^{2.5}}^\dagger$ & 	$ (4.68 \pm 0.95) \times 10^{-16} $ & $-$ \\
32$^{*}$ & VV1818Ori 				& Orion 		& $633.4\pm22.7 $	 $^a$ & $292_{18}^{19}$ &  $ (3.50 \pm 0.55) \times 10^{-13} $ & $58 \pm 4$ \\   
33 		 & 2MASSJ055749461405278 	& Orion 		& $466.0\pm44.6 $  $^a$   & $12.44_{1.96}^{1.97} $ &  $ (1.93 \pm 0.09) \times 10^{-14} $ & $-$ \\                        
34 		 & 2MASSJ055749181406080 	& Orion 		& $1039.5\pm334.4$  $^a$  & ${5.73_{0.93}^{1.17}}^\dagger$ &  $ (2.07 \pm 0.54) \times 10^{-15} $ & $-$ \\     
35$^{*}$ & WL16 					& Oph 			& $138.4 \pm 2.6$ $^d$	  & $15.05	_{1.28}^{1.13} $ &  $ (2.53 \pm 0.11) \times 10^{-13} $ & $3.7 \pm 0.2$ \\
36 		 & CG2010IRAS162882450W1 	& Oph 			& $138.4 \pm 2.6$ $^d$	  & ${0.009_{0.001}^{0.001}}^\dagger$ &  $ (8.25 \pm 1.30) \times 10^{-14} $ & $-$ \\      
37$^{*}$ & IRAS162894449 			& Sa 187 		& $350.3 \pm 2.5	$ $^e$& $60.5_{2.7}^{3.0} $ &  $ (1.24 \pm 0.19) \times 10^{-13} $ & $146 \pm 16$ \\
38 		 & JCMTSFJ1634294154700 	& L43 			& $160 \pm 16$       $^f$ & $3.63_{0.50}^{0.72} $ &  $ (1.31 \pm 0.54) \times 10^{-14} $ & $-$ \\      
39 		 & 2MASSJ164658260935197 	& L260 			& $322.09 \pm 9.7	$ $^e$& $3.15_{0.15}^{0.23} $ &  $ (5.38 \pm 1.05) \times 10^{-15} $ & $-$ \\
40 		 & IRAS182750040 			& Serpens 		& $383.1\pm12.2 $ $^a$	  & $16.2	_{1.0}^{1.1} $ &  $ (6.28 \pm 0.98) \times 10^{-14} $ & $-$ \\      
41$^{*}$ & 2MASSJ183646330110294 	& Serpens 		& $580.4\pm70.6 $  $^a$   & $63.8_{16.1}^{20.0}$ &  $ (8.39 \pm 3.75) \times 10^{-15} $ & $-$ \\
42 		 & VSCrA 					& CrA 			& $160.5\pm1.8 $  $^a$    & $7.13_{0.21}^{0.20} $ &  $ (9.83 \pm 0.45) \times 10^{-13} $ & $307 \pm 7$ \\      
43$^{*}$ & Parsamian21 				& Aquila 		& $400 \pm 100	$ $^g$ 	  & $9.71_{4.99}^{3.96} $ &  $ (1.01 \pm 0.30) \times 10^{-15} $ & $214 \pm 107$ \\      
44 		 & LDN1100 					& Cepheus 		& $561.8\pm107.8 $ $^a$   & ${26.52_{8.40}^{13.03}}^\dagger$ &  $ (4.80 \pm 0.92) \times 10^{-15} $ & $-$ \\         
45 		 & IRAS214455712 			& IC 1396 East 	& $360 \pm 36   $  $^h$   & $52.48_{14.67}^{23.23}$ &  $ (2.10 \pm 1.16) \times 10^{-15} $ & $-$ \\
46 		 & IRAS222666845 			& L1221, HH 363 & $200	\pm 20	$ $^i$ 	  & $2.61_{0.58}^{0.50} $ &  $ (4.87 \pm 0.76) \times 10^{-15} $ & $-$ \\
47 		 & IRAS222726358B 			& L1206 		& $950.0 \pm 95	$ $^i$ 	  & ${53.0_{9.7}^{8.7}}^\dagger$ &  $ (5.44 \pm 0.68) \times 10^{-14} $ & $-$ \\       
48 		 & EMLkHA233 				& Lacerta 		& $498.9 \pm  11.6$ 	  & $56.1_{3.6}^{2.1} $ &  $ (1.51 \pm 0.18) \times 10^{-14} $ & $-$ \\
49 		 & 2MASSJ230549766230011 	& Cep C  		& $1190 \pm 119$  $^h$	  & $2882_{729}^{455} $ &  $ (2.95 \pm 0.33) \times 10^{-14} $ & $-$ \\
50 		 & 2MASSJ000143254805189 	& Cepheus 		& $379.9\pm15.6 ^a$    & $4.01	_{0.39}^{0.55} $ &  $ (8.02 \pm 1.26) \times 10^{-15} $ & $-$ 
\enddata

\tablecomments{$^a$Parallax distance with {\it Gaia} EDR3 direct match \citep{GaiaCollaboration.Brown.ea2021}, Distance to the region (error is set to 10\% if not stated in literature): $^b$\cite{Krolikowski.Kraus.ea2021}, $^c$\cite{tob20}; $^d$\cite{ort18} $^e$\cite{Zari.Hashemi.ea2018}, $^f$\cite{Anglada.Rodriguez2002a}, $^g$\cite{Kospal.Abraham.ea2008}, $^h$\cite{Wouterloot.Brand1989}, $^i$\cite{Hilton.Lahulla1995}, $^j$assumed to be the same as FS TauA where {\it Gaia} EDR3 is available, $^k$distance estimated using photogeometric method from Gaia EDR3 data. \cite{bai20}. $^*$Class\,II or FUors. $^\dagger \lbol$ of YSOs for which no photometry at $\lambda > 100 \, \mu$m was available.}
\end{deluxetable*}

\begin{figure}
    \centering
    \includegraphics[width=\columnwidth]{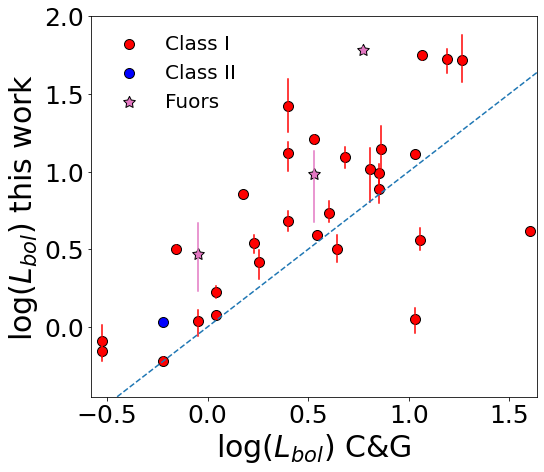}
    \caption{Comparison between the bolometric luminosity computed in this work with those computed by  CG10. FUors are included in this plot (pink stars). Class\,I and II are shown in red and blue circles, respectively. The uncertainty smaller than the symbol size is not plotted. The solid line indicates the one-to-one correspondence.}
    \label{fig:Lbol_comparison}
\end{figure}

We compared our results with the bolometric luminosity computed by  CG10. 
They provided results for only 34 sources among our sample by using the formula discussed in \citet[][Eq. 2]{connelley2007}, based on the IRAS fluxes (from 12\,$\mu$m to 100\,$\mu$m), and on the distance of the source, assuming the same SED model we adopt for $\lambda > 100$\,$\mu$m. 
Fig.\,\ref{fig:Lbol_comparison} shows the results of this comparison: 15 sources (44\%) have $\lbol$ values in agreement within $3\sigma$ by using the two methods; and for only three sources our $\lbol$ estimate is lower than those by CG10. 
We used different and updated distances than CG10, and we considered not only IRAS fluxes, but also more accurate and recent flux estimates at both shorter (e.g. {\it Gaia} 500\,nm) and longer (e.g. SCUBA 850\,$\mu$m) wavelengths. 
The different methodologies should be an indicator to not expect the same values.
However, we can not neglect the trend we see in Fig.\,\ref{fig:Lbol_comparison} that our $\lbol$ values are systematically larger than those provided by CG10. 
A possible interpretation of this trend is that integrating over a larger wavelength range, we include a larger portion of the emitted luminosity.
On the contrary, when our $\lbol$ estimates are lower, it is possible that these sources are affected by the overestimation already pointed out by  CG10. 
Moreover, the method adopted by  CG10 infer the sub-mm flux by fitting the same 36\,K blackbody to the IRAS flux at 100\,$\mu$m, while our analysis directly integrate observed fluxes, relying on similar model only when data at $\lambda > 100$\,$\mu$m are not available.
Looking at the SEDs in the Appendix\,\ref{app:sed}, we note that all the objects for which we computed smaller $\lbol$ estimates show particularly low fluxes for $\lambda > 100$\,$\mu$m,  which constrains the integral of the SEDs empirically, not considering any model.

In general, we are confident that our estimate of the bolometric luminosity are more robust because ($i$) we used the most updated distances based on {\it Gaia} data, ($ii$) we integrated for a larger wavelength range, and ($iii$) our SEDs are carefully and more densely sampled. All these improvements contribute to provide more accurate results.

\subsection{Accretion rates and stellar parameters}
\label{sect:acc_par}
Measuring the mass accretion rate in embedded protostars is challenging for two main reasons. 
First, Class\,I objects are not visible in the UV, where the Balmer jump directly traces the accretion luminosity. 
Secondly, the envelope contribution of Class\,I sources is entangled with the disk and the photosphere contributions, preventing us to compute the visual extinction ($A_V$), veiling, and stellar parameters with the standard methodology used for CTTS stars as spectral shape fitting and absorption features analysis.
For these reasons, we adopted a self-consisted method already used in the literature to analyse Class\,I protostars \citep{ant08, fio21}.

This procedure is based on the following assumptions: 
({\it i}) the bolometric luminosity is the sum of the stellar and accretion luminosity, $\lbol = \lstar + \lacc$, where the disk luminosity ($L_{\rm disk}$) is reprocessed during the accretion, thus $L_{\rm disk}$ is ``included" in the $\lacc$ contribution; ({\it ii}) envelope and disk contributions, which cannot be disentagled in Class\,I sources, can be described by the

\begin{figure}
    \centering
    \includegraphics[width=\columnwidth]{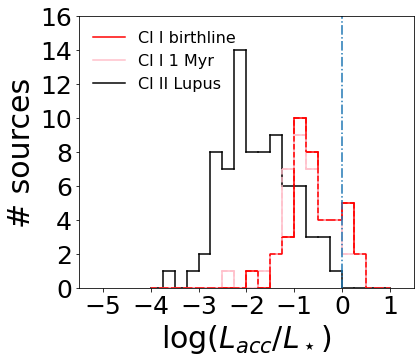}
    \includegraphics[width=\columnwidth]{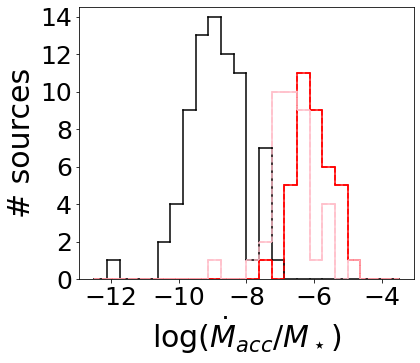}
    \caption{Histograms of the $\lacc/\lstar$ ratio (top panel) and $\macc/\mstar$ ratio (bottom panel) for Class I assuming the age of the birthline (red) and 1\,Myr (pink) compared to the histogram of the same quantity for the Class II sample of the Lupus star-forming region. The blue dashed-dotted line correspond to $\lacc/\lstar = 1$.}
    \label{fig:histo}
\end{figure}
\noindent veiling ($r_K$), therefore the absolute bolometric magnitude in $K$\,band is M$_{\rm bol} = \mbox{BC}_K + m_K + 2.5 \log(1 + r_K) - A_K - 5 \log(d/10 \mbox{pc})$; 
({\it iii}) the empirical relations between the luminosity of H{\footnotesize{I}} lines and the accretion luminosity found for CTTS are a good approximation for Class\,I stars \citep[][]{nis05a,fio21}, in particular the one regarding the $\brg$ line, $\log \lacc = a \log L_{Br\gamma} + b$, where $a = 1.19 \pm 0.10$ and $b = 4.02 \pm 0.51$ \citep{alc17}. 
Consequently, knowing the observed $K$\,band magnitude ($m_K$), $L_{Br\gamma}$, $\lbol$, the distance, and the $K$\,band veiling ($r_K$), and assuming the age of the object which sets the spectral type and thus the bolometric correction ($BC_K$), then the only free parameter is the extinction ($A_K$), which consistently provides the same $\lstar$ from
the computations in points ({\it i}) and ({\it ii}). 
For a detailed description of this method, we refer the reader to Sect.\,5.1 of \citet{fio21}.

To use this self-consistent method, we measured the flux of the $\brg$ line ($F_{Br\gamma}$) as follows. 
From the $K$\,band magnitude listed in Tab.\,\ref{tab:sample}, we computed the continuum flux that we used to convert the equivalent with of the $\brg$ line ($W_{eq}$) into flux. Errors were computed by propagating the uncertainties provided by \citet{con10}, see Tab.\,\ref{tab:results}. 
Then, the line luminosity is $L_{Br\gamma} = 4 \pi d^2 F_{Br\gamma}$. 
We used $\lbol$ computed as described in Sect.\,\ref{sect:lbol}, and the veiling from  CG10. 
We assumed that our sources are
located between the birthline \citep[as defined by][]{pal93}, and the 1\,Myr isochrone of the \citet{sie00} models. 
Using the self-consistent method described above, we obtained both the stellar and the accretion luminosity for each protostar.

We determined the stellar mass and radius ($\mstar$, $\rstar$) using both the birthline and the 1\,Myr isochrone from \citet{sie00}; and the mass accretion rate using the relation:
\begin{equation}
\macc \sim \left( 1 - \frac{\rstar}{R_{in}} \right)^{-1} \frac{\lacc \rstar}{G \mstar}
\label{eq:macc}
\end{equation}
where $R_{in}$ is the inner-disk radius which we assume to be $R_{in} \sim 5 \rstar$ \citep{har98}, and $G$ is the gravitational constant. 
The average errors on the accretion luminosity, stellar radius and mass are 0.4, 0.6, and 0.1\,dex, which result in a cumulative error for the mass accretion rate of 0.8\,dex \citep[][]{fio21}. 
To these uncertainties we should also add the uncertainty due to the sources variability, which is estimated to be about 0.5\,mag in flux \citep[][]{lorenzetti2013}, propagating a variation on the flux of about 50\%.
The accretion luminosity, the mass accretion rate, and the stellar parameters of our sample ranging from the birthline to 1\,Myr of age are listed in Tab.\,\ref{tab:results}. 

Fig.\,\ref{fig:histo} shows histograms of $\lacc/\lstar$ (top panel) and $\macc/\mstar$ (bottom panel) ratios of our Class\,I sample (red line if we assume sources on the birthline, pink line if we assume the age of the sources is 1\,Myr) compared to the Lupus Class\,II sample \citep[black line,][]{manara2022pp7}. We chose Class\,II objects in this cloud because they are representative for Class\,II PMS stars, and well studied in the recent past \citep[][]{alc14, alc17}.

The histogram of our Class\,I sample peaks at higher values ($\log(\lacc/\lstar) \sim -0.9$ and $\log(\macc/\mstar/yr) \sim -6.5$) compared to Class\,II sample ($\log(\lacc/\lstar) \sim -2.2$ and $\log(\macc/\mstar/yr) \sim -9.0$) for both the $\lacc/\lstar$ and $\macc/\mstar$ ratios. This suggests that the accretion in Class\,I objects is more intense than during the Class\,II stage.
We performed a two-sample Kolmogorov-Smirnov test (KS test) on the Class\,I and Lupus Class\,II populations for both the $\lacc/\lstar$ and $\macc/\mstar$ distributions. 
We obtained a probability of 0.32 for the Class\,I on the birthline and 0.35 for the Class\,I 1\,Myr old for the luminosity. This means that the probability for the Class\,I and the Class\,II $\lacc/\lstar$ values to be drawn from the same statistical distribution is about 1/3: $0.32 < p <0.35$.
The same analysis on the $\macc/\mstar$ distributions results in a probability of 0.10 for the Class\,I on the birthline and 0.15 for the Class\,I 1\,Myr old. In other words, the probability for the Class\,I and the Class\,II $\macc/\mstar$ values to be drawn from the same statistical distribution is particularly low: $0.10 < p <0.15$.
While the $\lacc/\lstar$ distributions largely overlap each other (more than half of the Class\,I histograms are inside the Class II one), in the case of $\macc/\mstar$ distribution, only their tail-ends overlap. 
This difference corresponds to the different KS test results, showing that the $\lacc/\lstar$ distribution between Class\, I and II is more similar than the $\macc/\mstar$ distribution. 
This can be explained by the presence of the stellar parameters in Eq.\,\ref{eq:macc}, in particular the radius, which is larger for younger sources in the evolutionary models we used. 
Also, it is important to keep in mind that while the Class\,II is a complete sample from Lupus, the Class\,I sample is not complete, being composed only of the brightest sources, from different star forming regions (SFRs). 
Thus, any intrinsic differences between the SFRs can blur the differences between Classes by widening the FWHM of either histogram. 
This means that we mostly analyze the left-side of the full histogram for all the Class\,I YSOs. 
In turn, because the p-value in the KS test uses the mean of the distributions, it is possible to speculate that we obtain lower probabilities than considering also the most embedded Class\,I YSOs. 
More data of a complete sample belonging to a single SFR are needed to draw firm conclusions.
However, the fact that we can already see differences in the $\macc/\mstar$ histograms with our incomplete and biased sample indicates that the differences will probably be much more pronounced, once the Class\,I sample is complete.

It is also worth to note that only for 8 Class\,I on the birthline and for 7 Class\,I on the 1\,Myr evolutive track the accretion luminosity is larger than its stellar luminosity ($\lacc/\lstar > 1$), see dot-dashed cyan vertical line in Fig.\,\ref{fig:histo}. This could also be due to an observational bias, since we are studying the less embedded Class\,I young stars.

\begin{figure}[t]
    \centering
    \includegraphics[width=0.5\textwidth]{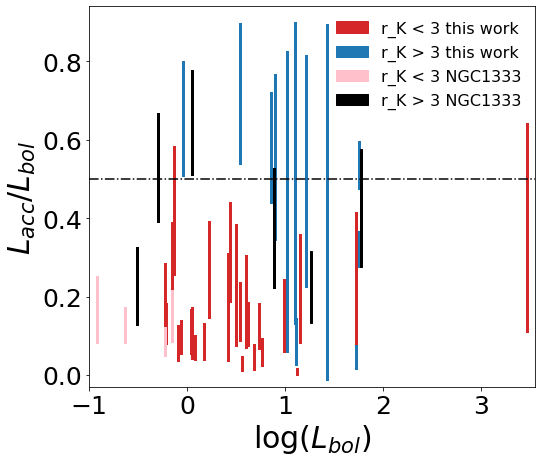}
    \caption{The $\lacc/\lbol$ ratio as a function of the bolometric luminosity for Class\,I with veiling smaller than three (red) and higher than three (blue). We compare our results with Class\,I in Perseus NGC1333 cluster for low-veiled (pink) and high-veiled (black) sources.
    } 
    \label{fig:LaccLbolratio}
\end{figure}

To investigate which is the dominant component of the luminosity of our protostars, we plot in Fig.\,\ref{fig:LaccLbolratio} the accretion fraction as a function of the bolometric luminosity (red and blue lines represent low-veiling sources, i.e. $r_K < 3$, and high-veiling sources, i.e. $r_K > 3$, respectively), and we compare this sample with the NGC\,1333 protostars from \citet[][]{fio21} (pink and black lines represent low-veiled and high-veiled sources, respectively). 
Most of the sources in their main accretion phase (i.e. $\lacc/\lbol > 0.5$) have high veiling value and only 2 of them have low veiling value, showing that the accretion luminosity and the veiling are related, as already noted in previous works \citep[e.g.][]{cal04, fis11, fio21}. 

We note that only for 15 out of 39 young stars (38\%) the accretion luminosity has value compatible with the condition of being the main contribution to the bolometric luminosity, even if $\lacc/\lstar \gg 1$ is what is expected for Class\,I protostars in general.
We think this result shows the above mentioned bias in our sample. 
This strongly suggests that our conclusions are not meant to be intended for Class\,I objects in general, but only for the less embedded Class\,I YSOs for which it was possible to estimate the veiling.

\subsection{Disk dust masses}
\label{sect:analysis_disk_mass}

For the overall sample described in Sect.\,\ref{sect:sample} we searched for (sub)millimeter fluxes, finding suitable observations for only 60 of them. 
We performed a coordinate and sources name search across the literature and the archival interferometric data. 
We included a dust mass measurement in our analysis if the flux measurement was available at $< 1"$ resolution to mitigate the possible contribution from the envelope. 
In the sub-arcsecond regime with size of the beam comparable to the disk size the envelope contribution is usually negligible, especially for the Class\,I systems where the envelope is largely dissipated \citep{tyc20}. 
We collected data on dust disk fluxes for 21 sources from Tab.\,\ref{tab:sample}, for 12 sources from Tab.\,\ref{tab:clI_literature}, and for all the sources in Tab.\,\ref{tab:indirect_sources}. Tab\,\ref{tab:submm_fluxes} list mm-fluxes and references for all the sources.

From the flux density ($F_\nu$) we calculated the dust mass by inverting the modified black-body equation:
\begin{equation}
\label{eq:dustmass}
M_{\rm dust}=\frac{d^2F_\nu}{{\kappa}_\nu(\beta) B_\nu(T_{\rm dust})}\,
,\end{equation}

\noindent where $d$ is the distance to the source, $B_\nu$ is the Planck function for the dust temperature $T_{\rm dust}$, and $\kappa_\nu$ is the dust opacity at the frequency of the observation $\nu$.
The equation is accurate for optically thin emission, otherwise it provides a lower limit on the dust mass measurement. 

A dust temperature value of 30\,K is assumed, as typically assumed for embedded young stars \citep[i.e.][]{ans16}. 
If the temperature is lower, similarly to Class\,II disks, the total dust mass would be higher.
We adopted a dust opacity value of 0.00899\,g\,cm$^{-2}$ \citep{oss94}, and a spectral emissivity slope $\beta=1$. 
With uniform assumptions on dust properties we are not introducing additional discrepancy between the disks measured within different observing projects. 

We are aware that there is no agreement on the accuracy of the disk mass estimation \citep[see the on-going debate on][and references therein]{miotello2022pp7, manara2022pp7}. 
In brief, some studies suggest a severe underestimation of the disk mass because of the optical thickness or dust scattering \citep{Zhu.Zhang.ea2019}. 
On the contrary, \cite{Sheehan.Tobin.ea2022} show that protostellar dust disk masses can be overestimated by using the isothermal disk assumption, and this effect is higher the less massive are the disks. 
These two opposite effects would increase the disk masses spread. In other words, more massive disks would be even more massive while the low-mass end would be even less massive.

\begin{figure}[b]
    \centering

        \includegraphics[width=\columnwidth]{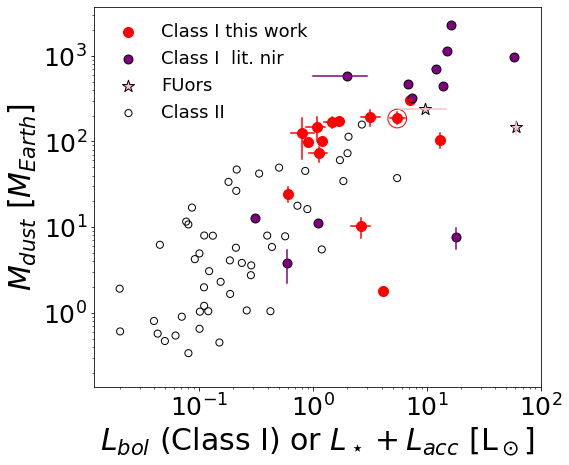}

    \caption{Dust disk mass as a function of bolometric luminosity for the sample of Class\,I sources in Tabs.\,\ref{tab:clI_literature} (purple dots), \ref{tab:first_res} (red dots), and for FUors in Tab.\,\ref{tab:sample} that have disk estimate. For Class\,II, the disk mass is plotted against the sum of stellar and accretion luminosity. The big empty red circle highlights source ID 02, for which no photometry at $\lambda > 100 \, \mu$m was available and therefore $\lbol$ is probably underestimated. Other sources with the same issue are not shown in this plot since we do not provide $\mdust$ estimates for them.}
    \label{fig:lbol_vs_diskmass}
\end{figure}

Fig. \ref{fig:lbol_vs_diskmass} presents disk masses calculated for our sample of Class\,I (red dots) and FUors (pink stars), for other Class\,I present in the literature (purple dots), and for Class\,II of Lupus (empty circles).
Class\,I young stars follow the same trend as Class\,II, extending the Class\,II linear distribution on the upper-right part of the plot, which shows that $\mdisk$ is higher in Class\,I than in Class\,II. 
We note that FUors population seems to deviate from this general trend. 
\citet{Kospal.CruzSaenzdeMiera.ea2021} found that FUors have larger disk masses than either regular Class\,I or Class\,II objects. 
A possible explanation for this might be that FUor disks are also smaller and more optically thick, so by using the optically thin assumption we did (by converting the continuum fluxes to disk masses), we underestimated the dust mass in FUors more severely than for regular disks. 
However, the significance of this deviation can not be established with only three sources and further data and more accurate analisys are needed to confirm the hint suggested by this figure.

\startlongtable
\begin{deluxetable*}{ccccccccc}
\tablecaption{Stellar and accretion parameters of the 39 Class\,I from \citet[][]{con10} computed in this work.\label{tab:results}}
\tablewidth{0pt}
\tablehead{
\colhead{ID} & age & $A_ V$ & $\lacc$ & $\lstar$ & $T_{eff}$ & $\mstar$ & $\rstar$ & $\macc$ \\
\colhead{  } &     & mag & $\lsun$ & $\lsun$ & K         &$\msun$ & $\rsun$ & $\msun/$yr}
\decimalcolnumbers
\startdata
01 & bl - 1Myr & $18.61 - 15.95$ & $ 0.14 -  0.10$ & $  1.36 - 1.40$ & $ 3126 - 3487$ & $0.22 - 0.35$ & $4.00 - 3.05$ & $-7.03 / -7.49$ \\
02 & bl - 1Myr & $29.21 - 28.02$ & $ 0.71 -  0.61$ & $  4.87 - 4.97$ & $ 3673 - 3958$ & $0.47 - 0.68$ & $5.35 - 4.43$ & $-6.52 / -6.83$ \\
03 & bl - 1Myr & $23.15 - 21.94$ & $ 0.23 -  0.20$ & $  4.64 - 4.67$ & $ 3652 - 3935$ & $0.45 - 0.65$ & $5.31 - 4.42$ & $-7.06 / -7.36$ \\
05 & bl - 1Myr & $15.69 - 13.61$ & $ 0.49 -  0.39$ & $  1.19 - 1.29$ & $ 3073 - 3467$ & $0.21 - 0.34$ & $3.75 - 3.04$ & $-6.48 / -6.90$ \\
06 & bl - 1Myr & $ 9.93 -  7.71$ & $ 0.09 -  0.07$ & $  1.11 - 1.13$ & $ 3090 - 3407$ & $0.21 - 0.32$ & $3.72 - 2.96$ & $-7.23 / -7.63$ \\
07 & bl - 1Myr & $41.51 - 39.76$ & $ 0.13 -  0.10$ & $  0.95 - 0.98$ & $ 3020 - 3313$ & $0.20 - 0.29$ & $3.50 - 2.77$ & $-7.08 / -7.44$ \\
09 & bl - 1Myr & $20.39 - 19.99$ & $ 0.32 -  0.31$ & $  0.43 - 0.44$ & $ 2834 - 2936$ & $0.15 - 0.17$ & $3.00 - 2.35$ & $-6.62 / -6.78$ \\
10 & bl - 1Myr & $21.65 - 19.50$ & $ 0.58 -  0.44$ & $  3.54 - 3.68$ & $ 3448 - 3868$ & $0.34 - 0.58$ & $4.82 - 4.11$ & $-6.50 / -6.92$ \\
11 & bl - 1Myr & $43.04 - 43.04$ & $ 0.60 -  0.59$ & $  0.29 - 0.30$ & $ 2818 - 2835$ & $0.15 - 0.14$ & $3.01 - 2.26$ & $-6.32 / -6.41$ \\
13 & bl - 1Myr & $27.45 - 26.68$ & $ 0.17 -  0.16$ & $  0.55 - 0.56$ & $ 2851 - 3040$ & $0.16 - 0.20$ & $3.08 - 2.51$ & $-6.97 / -7.19$ \\
15 & bl - 1Myr & $28.29 - 26.67$ & $ 0.07 -  0.05$ & $  0.77 - 0.78$ & $ 2917 - 3200$ & $0.17 - 0.25$ & $3.14 - 2.64$ & $-7.36 / -7.67$ \\
16 & bl - 1Myr & $ 2.63 -  1.99$ & $ 2.54 -  2.44$ & $  0.90 - 1.01$ & $ 2972 - 3264$ & $0.19 - 0.28$ & $3.50 - 2.82$ & $-5.72 / -6.00$ \\
17 & bl - 1Myr & $36.77 - 35.70$ & $ 0.08 -  0.07$ & $  0.53 - 0.54$ & $ 2834 - 3037$ & $0.15 - 0.20$ & $3.00 - 2.48$ & $-7.23 / -7.49$ \\
18 & bl - 1Myr & $ 8.96 -  6.78$ & $ 0.62 -  0.48$ & $  2.89 - 3.03$ & $ 3388 - 3780$ & $0.31 - 0.51$ & $4.67 - 3.84$ & $-6.46 / -6.88$ \\
20 & bl - 1Myr & $27.81 - 26.81$ & $ 4.54 -  3.98$ & $  3.25 - 3.80$ & $ 3413 - 3849$ & $0.34 - 0.59$ & $4.77 - 4.06$ & $-5.60 / -5.97$ \\
21 & bl - 1Myr & $30.49 - 26.48$ & $ 3.41 -  2.11$ & $ 11.39 -12.69$ & $ 3960 - 5029$ & $0.79 - 2.37$ & $6.45 - 4.38$ & $-5.99 / -6.84$ \\
22 & bl - 1Myr & $15.20 - 10.33$ & $ 1.25 -  0.70$ & $ 11.72 -12.27$ & $ 4050 - 5158$ & $0.92 - 2.67$ & $6.86 - 4.03$ & $-6.48 / -7.44$ \\
23 & bl - 1Myr & $ 6.51 -  4.12$ & $ 1.52 -  1.09$ & $  8.26 - 8.70$ & $ 3890 - 4722$ & $0.68 - 1.91$ & $6.07 - 4.21$ & $-6.33 / -7.04$ \\
24 & bl - 1Myr & $ 4.88 -  3.58$ & $ 0.09 -  0.08$ & $  0.79 - 0.80$ & $ 2951 - 3217$ & $0.18 - 0.25$ & $3.29 - 2.63$ & $-7.22 / -7.53$ \\
25 & bl - 1Myr & $ 3.97 -  2.62$ & $ 0.89 -  0.75$ & $  1.88 - 2.02$ & $ 3236 - 3630$ & $0.26 - 0.42$ & $4.45 - 3.41$ & $-6.24 / -6.64$ \\
26 & bl - 1Myr & $12.89 - 11.04$ & $ 4.90 -  4.30$ & $  5.36 - 5.96$ & $ 3563 - 4187$ & $0.49 - 1.14$ & $5.31 - 4.20$ & $-5.74 / -6.24$ \\
27 & bl - 1Myr & $ 4.72 -  3.75$ & $ 0.10 -  0.09$ & $  0.55 - 0.57$ & $ 2919 - 3051$ & $0.17 - 0.20$ & $3.26 - 2.48$ & $-7.20 / -7.39$ \\
28 & bl - 1Myr & $17.36 - 15.57$ & $ 0.13 -  0.10$ & $  0.99 - 1.02$ & $ 3020 - 3331$ & $0.20 - 0.29$ & $3.50 - 2.86$ & $-7.09 / -7.44$ \\
29 & bl - 1Myr & $23.28 - 21.00$ & $ 0.59 -  0.45$ & $  2.07 - 2.21$ & $ 3255 - 3652$ & $0.26 - 0.43$ & $4.46 - 3.46$ & $-6.43 / -6.87$ \\
30 & bl - 1Myr & $16.21 - 14.04$ & $ 0.78 -  0.61$ & $  2.48 - 2.65$ & $ 3312 - 3715$ & $0.28 - 0.47$ & $4.56 - 3.66$ & $-6.38 / -6.81$ \\
31 & bl - 1Myr & $32.89 - 27.37$ & $ 0.14 -  0.07$ & $ 12.71 -12.78$ & $ 4027 - 5188$ & $0.87 - 2.73$ & $6.71 - 4.19$ & $-7.42 / -8.41$ \\
33 & bl - 1Myr & $20.70 - 18.35$ & $ 6.99 -  6.05$ & $  5.46 - 6.40$ & $ 3548 - 4371$ & $0.48 - 1.50$ & $5.28 - 3.68$ & $-5.54 / -6.23$ \\
34 & bl - 1Myr & $18.50 - 17.15$ & $ 0.35 -  0.30$ & $  5.50 - 5.55$ & $ 3716 - 4028$ & $0.50 - 0.76$ & $5.43 - 4.57$ & $-6.88 / -7.20$ \\
38 & bl - 1Myr & $17.66 - 15.55$ & $ 0.12 -  0.09$ & $  3.62 - 3.65$ & $ 3509 - 3846$ & $0.37 - 0.56$ & $4.96 - 4.05$ & $-7.25 / -7.64$ \\
39 & bl - 1Myr & $28.21 - 26.36$ & $ 0.77 -  0.63$ & $  2.42 - 2.56$ & $ 3349 - 3715$ & $0.30 - 0.47$ & $4.64 - 3.66$ & $-6.37 / -6.77$ \\
40 & bl - 1Myr & $21.99 - 19.73$ & $ 9.53 -  8.01$ & $  6.64 - 8.16$ & $ 3684 - 4517$ & $0.55 - 1.65$ & $5.53 - 4.12$ & $-5.41 / -6.05$ \\
42 & bl - 1Myr & $ 6.77 -  5.79$ & $ 4.34 -  3.99$ & $  2.78 - 3.14$ & $ 3410 - 3781$ & $0.33 - 0.52$ & $4.79 - 3.86$ & $-5.60 / -5.93$ \\
44 & bl - 1Myr & $40.07 - 36.23$ & $11.83 -  9.94$ & $ 16.99 -18.89$ & $ 3940 - 5088$ & $0.79 - 2.36$ & $6.42 - 4.86$ & $-5.75 / -6.50$ \\
45 & bl - 1Myr & $53.39 - 46.30$ & $ 7.65 -  3.35$ & $ 49.12 -53.42$ & $ 4074 - 5754$ & $0.96 - 3.06$ & $7.02 - 5.69$ & $-5.76 / -6.72$ \\
46 & bl - 1Myr & $30.71 - 28.37$ & $ 0.31 -  0.23$ & $  2.26 - 2.33$ & $ 3312 - 3673$ & $0.28 - 0.45$ & $4.56 - 3.52$ & $-6.77 / -7.20$ \\
47 & bl - 1Myr & $ 6.59 -  3.61$ & $13.27 -  8.05$ & $ 39.26 -44.48$ & $ 4074 - 5754$ & $0.96 - 3.06$ & $7.02 - 5.69$ & $-5.52 / -6.27$ \\
48 & bl - 1Myr & $40.58 - 36.44$ & $29.99 - 18.03$ & $ 25.29 -37.26$ & $ 4074 - 5754$ & $0.96 - 3.06$ & $7.02 - 5.69$ & $-5.06 / -5.88$ \\
49 & bl - 1Myr & $47.24 - 42.53$ & $1263  -   797$ & $1482   -1948 $ & $ 4074 - 5754$ & $0.96 - 3.06$ & $7.02 - 5.69$ & $-3.48 / -4.31$ \\
50 & bl - 1Myr & $20.17 - 18.40$ & $ 0.79 -  0.64$ & $  3.30 - 3.45$ & $ 3448 - 3802$ & $0.34 - 0.53$ & $4.82 - 3.91$ & $-6.42 / -6.80$ 
\enddata
\end{deluxetable*}

\section{Discussion}

\begin{figure*}
    \centering
    \includegraphics[width=1.05\columnwidth]{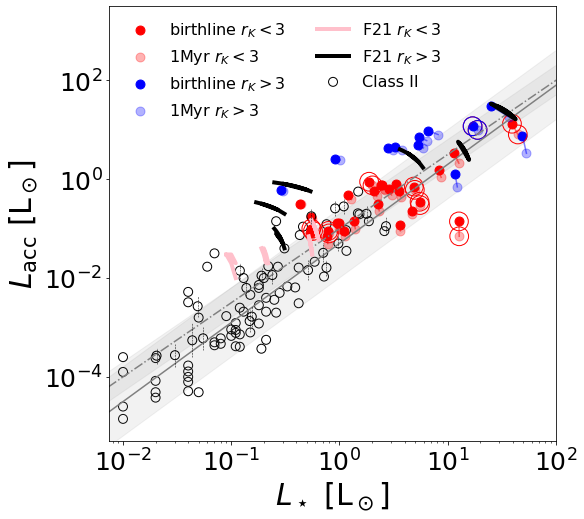}
    \includegraphics[width=1.05\columnwidth]{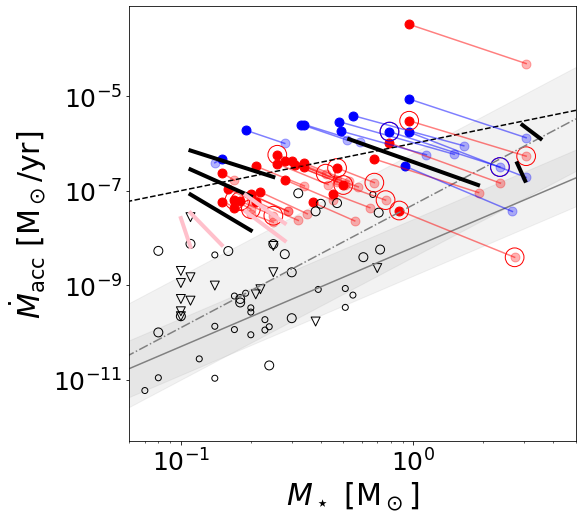}
    \caption{{\it Left}: Accretion luminosity $vs.$ stellar luminosity. Red and blue dots are Class\,I with $r_K < 3$ and $r_K > 3$, respectively, and are blurred depending on the assumed age as described in the legend. Source for which photometry at $\lambda > 100 \, \mu$m is not available are surrounded by big red/blue empty circles. The two values (birthline and 1\,Myr) are linked by a line to show that all the combination of accretion and stellar parameters between the two dots are possible. Similarly, the ranges of possible parameters assuming the age between the birthline and 1\,Myr are shown for NGC\,1333 cluster depending on the veiling, pink for $r_K > 3 $ and black for $r_K > 3$. Empty black circles are Class\,II of Lupus and NGC1333 Perseus clouds, and triangles showed the related upper limits. The solid and point-dashed lines are the best fit of the Lupus and NGC1333 Class\,II, respectively \citep[][]{fio21}. The grey region corresponds to the standard deviation of the fits. {\it Right}: Mass accretion rate $vs.$ stellar mass. All the symbols are as in the left panel. The dash-dotted black line show $\log \macc \propto \log \mstar$.}
    \label{fig:LaccLstar}
\end{figure*}

\subsection{Accretion properties versus stellar properties}

The relations between the accretion and stellar luminosity, and the mass accretion rate and the stellar mass, are well established for CTTS \citep[see][and references therein]{manara2022pp7}. 
The same relations have been investigated recently for Class\,I YSOs by \citet{fio21}. They show that Class\,I of the NGC1333 cluster have higher $\macc$ than Class\,II of the same cluster with similar stellar mass.
While this result is demonstrated for the NGC1333 cluster, further observations are needed to see whether this conclusion is valid for other star-forming regions and/or in general for Class\,I objects.

Fig.\,\ref{fig:LaccLstar} (left panel) shows the accretion luminosity as a function of the stellar luminosity for our sample of Class\,I (red and blue dots depending on the veiling) and for Class\,I of NGC1333 (pink and black segments depending on the veiling). 
We plot also the Class\,II samples of Lupus and NGC1333 (empty circles), as a comparison.
The accretion luminosity in our sample ranges between $0.03$\,$\lsun$ and $1263$\,$\lsun$ for stellar luminosity between 0.28\,$\lsun$ and 3.65\,$\lsun$.
This figure shows that the less veiled sources (red filled circles) are compatible with the Class\,II trend, while the most veiled sources are the ones which deviate the most from the Class\,II fit to higher values of $\lacc$ for $\lstar < 1$\,$\lsun$. 
This veiling-dependent trend is valid also for the NGC1333 Class\,I population (pink and black lines in the plot).
We note that the Class\,II NGC1333 fit (grey point-dashed line) reproduces better than the Class\,II Lupus fit (grey solid line) the low-veiled Class\,I distribution, while the high veiled Class\,I objects represent the sub-sample with the wider spread in this plot.

We also plotted the $\macc \, vs. \, \mstar$ distribution in Fig.\,\ref{fig:LaccLstar} (right panel). 
The mass accretion rate of our sample ranges between  $\log (\macc/\msun/{\rm yr}) = -8.41$ and $-3.48$, for stellar masses between 0.13\,$\msun$ and 3.06\,$\msun$.
In this plot, the Class\,I sample is not in agreement with the Class\,II trend, showing higher accretion values than the Class\,II with similar stellar mass. 
The most veiled sources are accreting more than less veiled YSOs with similar mass, with the exception of two stars more massive than 1\,$\msun$.
We note that for both the $\lacc \, {\rm vs.} \,  \lstar$ and the $\macc \, {\rm vs.} \,  \mstar$ distributions, the only source which lies below the Class\,II YSOs trend (grey band) is a source for which photometry at $\lambda > 100 \, \mu$m was not available. Therefore, in this case, it should be considered that the bolometric luminosity we provide is underestimated, and the accretion luminosity is potentially higher.

The assumption on the age results in a large uncertainty (displayed in the plots by the line which links the birthline and 1\,Myr results for each source), in particular in $\mstar$. This prevents us from fitting the Class\,I YSOs $\macc \, {\rm vs.} \, \mstar$ distribution. 
We plot a linear relation with a unitary slope (black dashed line) to guide the reader observing that the Class\,I sample seems to present a flatter distribution with respect to the Class\,II sources, whose slope is about 2 \citep[$2.1 \pm 0.2$ in particular for Lupus,][]{fio21}. 
We made some fitting attempts 
considering for the overall sample the same age. 
Assuming that all the sources are on the birthline, we obtained $\log \macc = (1.64 \pm 0.40) \log \mstar + (-5.63 \pm 0.21)$, and by assuming they are all 1\,Myr old, we found $\log \macc = (0.59 \pm 0.27) \log \mstar + (-6.64 \pm 0.12)$. 
We stress that these fits are not reliable, since we do not know the exact age of each source, but can be seen as possible extremes, considering that at least some objects are probably younger. 
It is interesting that by assuming the oldest age for the overall sample, the intercept value is $-6.64 \pm 0.12$, larger than for Lupus \citep[$-8.2 \pm 0.10$,][]{fio21} and compatible within the error with the intercept of the NGC1333 sample \citep[$-7.3 \pm 0.6$,][]{fio21}, whose age is estimated to be $\sim 1$\,Myr. 

The mass accretion rate is predicted to correlate with the disk mass. This is predicted by viscous models \citep[e.g.,][]{har98,lodato17,rosotti17} and MHD wind models \citep[e.g.,][]{Tabone.Rosotti.ea2021}. This correlation was indeed observed for Class\,II YSOs \citep[][]{man16}. 
The data presented here allow a similar exploration of this relation in the Class\,I phase on a statistical significant sample. 
We refer the reader to the dedicated letter \citep[][]{fiorellino2022letter} where we studied the $\macc \, \mbox{vs.} \, \mdisk$ distribution, comparing Class\,I and Class\,II samples.

\subsection{Stellar Mass vs. Disk Mass} \label{sect:starsvsdisk}

\begin{figure}
    \centering
        \includegraphics[width=\columnwidth]{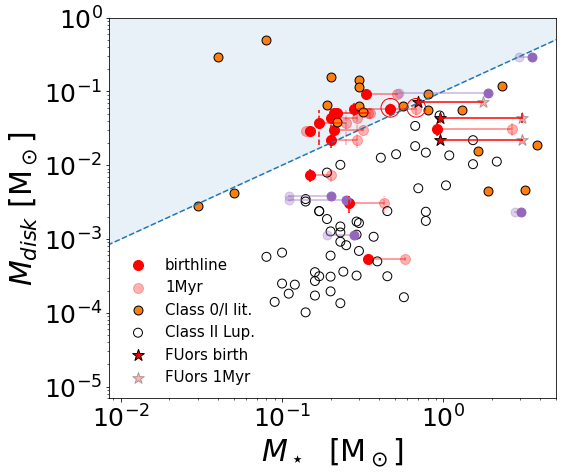}
 
    \caption{Disk mass versus stellar mass. Red dots are our Class\,I results assuming they are on the birthline, the pink dots are the same sources results assuming they are 1\,Myr old. Big empty red circles highlight source ID 02, for which no photometry at $\lambda > 100 \, \mu$m was available. Other sources with the same issue are not shown in this plot since we do not provide $\mdust$ estimates for them. Purple dots are sources in Tab.\,\ref{tab:clI_literature}, where blurring indicates the age for the NGC1333 sources, as described above. Orange dots are sources in Tab.\,\ref{tab:indirect_sources}. 
    Red stars and FUors from App.\ref{app:fuors}. 
    The blue dashed line marks the disk instability regime M$_{\rm disk} > 0.1\,\mstar$ region, filled in light blue.}
    \label{fig:mdisk_mstar}
\end{figure}

In Fig. \ref{fig:mdisk_mstar} we show the stellar mass of the protostars compared with their disk mass. 
Class\,I have systematically larger disk mass than Lupus Class\,II stars, as expected by the evolutionary path.

The $\mdisk \, vs. \, \mstar$ distribution of Class\,I samples is flat compared with steeper slope for Class\,II systems. 
The evolution of this trend for Class\,II has been studied in previous works \citep[e.g.][]{pas16, ans17, testi2022}, concluding that the slope becomes steeper with time. Therefore, our finding is consistent with this trend.

Models of early disk formation and evolution suggest rapid change of the $\mdisk \, vs. \, \mstar$ relation in the first 0.1\,Myr of system evolution \citep{Hennebelle.Commercon.ea2020}. 
Therefore the low ratio seen for only young sources is consistent and suggest extreme youth of those systems.

An open question regarding accretion is how much the episodic accretion is statistically important in the star formation process, since the eruptive objects list is nowadays confined to about 50 sources on different evolutionary stages \citep[][]{fisher2022pp7}. 
According to the current eruptive accretion scenario, the disk instability is fundamental to trigger the extremely strong outbutsts of FUors and EXors. Therefore, it makes sense to look for disk instability hints in ``steady" accretors. Signatures of disk fragmentation and other effects of instability in the young disks have been observed in several works \citep[e.g.,][]{Tobin.Kratter.ea2016,Alves.Caselli.ea2019}.
Thanks to the sensitive observations, new techniques can even trace a past outburst studying the ice-line radius or the presence of outbursts tracers \citep[see][and references therein]{fisher2022pp7}.
Indeed, \citet[][]{Kospal.CruzSaenzdeMiera.ea2021} found that about $2/3$ of FUors in their study may have gravitationally unstable disk, which can cause the typical strong outbursts. 

Motivated by this, we checkd in Fig.\,\ref{fig:mdisk_mstar} the stability of the disks in our Class\,I sample (including some FUors). 
We plot a blue dashed line which represents the edge of the instability disk regime (light-blue region), where $\mdisk > 0.1 \mstar$ \citep[Eq.\,3 in][]{Kratter.Lodato2016}.  
The plot shows that 81\% (17 out of 21) of the sources in our sample, 75\% (9 out of 12) of sources in Tab.\,\ref{tab:clI_literature} (purple dots), and 78\% (21 out of 27) of sources in Tab.\,\ref{tab:indirect_sources} (orange dots) lie in the light-blue region. 
This means that while all Class\,II disks appear stable, the majority of protostars shows evidence of  gravitationally unstable disks, suggesting that these sources experience some strong outbursts at some point during their evolution. 

We note that all the FUors do not lie in the instability region, as supposed for eruptive sources.  We think that this is because we analysed our FUors assuming their disks were optically thin, underestimating their disk mass. The results in this paper confirm that this method is not suitable to determine FUors parameters.

Since the sources in our sample are known to have gravitational stable disks, we looked for past or future outbursts signatures. 
We checked whether our sources were analyzed in \citet{Hsieh2019}, where they looked for N$_2$H$^+$ as a tracer of recent outburst. 
Among our sources in Tab.\,\ref{tab:sample}, only 2MASSJ03331284$+$3121241 has been studied. 
They found no clear detection for this star, thus they suggested that the envelope has pretty much dissipated. 
Because it lies in the instability region of Fig.\,\ref{fig:mdisk_mstar}, we can speculate this source will experience a future outburst. Concerning young stars in  Tabs.\,\ref{tab:clI_literature} and \ref{tab:indirect_sources}, four objects were observed by \citet{Hsieh2019}: 2MASSJ03283968$+$3117321,  which shows very weak or no emission and therefore its envelope may have also dissipated; 2MASSJ03285842$+$3122175 and 2MASSJ03290149$+$3120208, both stable according to our analysis; and V512\,Per, whose $\mdisk/\mstar$ ratio suggests it is stable, but it has been suggested that it is currently in outburst \citep[][]{Eisloeffel1991}. 
This would mean that V512\,Per is one of the few young stars in outburst within the stability 
region \citep[][]{Kospal.CruzSaenzdeMiera.ea2021} and thus its outburst must have been triggered by another physical process such as an interaction with its 
companion or a fly-by event \citep[][]{vorobyov2021}. 
Another possible explanation is that since we use the optically thin assumption, the disk mass is underestimated. In this case, a higher mass may shift it to the unstable region.
However, our results show that Class\,I disks, as FUors, have high fraction of gravitationally unstable disks. 

\section{Conclusions}
We analyzed available data for a sample of Class\,I protostars composed by 50 YSOs from \citet[][see Tab.\,\ref{tab:sample}]{con10}, for which we computed an accurate estimate of the bolometric luminosity (see Tab.\,\ref{tab:first_res}) and, for the first time, we provide the accretion and stellar parameters for 39 among them (Tab.\,\ref{tab:results}).
We also select 18 YSOs already analyzed with very similar methodologies than we used in this work (see Tab.\,\ref{tab:clI_literature}), and 27 YSOs studied with different methods (see Tab.\,\ref{tab:indirect_sources}). 
For the overall sample, we computed the disk dust mass, when archival interferometric data were available.
In this way, we built the largest sample of accretion, disks, and stellar properties ever analyzed for protostars (Class\,0 and I).   
We briefly summarized the key conclusions of this work:
\newline $-$ The accretion luminosity is higher in Class\,I than in Class\,II YSOs. This effect is smaller for the $\lacc \, vs. \, \lstar$ distribution of the low-veiled ($r_K < 3$) Class\,I objects. 
This highlighs the crucial need of accurate veiling estimates for the accretion study of embedded YSOs 
\newline $-$ The mass accretion rate in Class\,I sources is systematically higher than for Class\,II. Although the uncertainty on the age prevents us from providing reliable fit of our data, the $\macc \, {\rm vs.} \, \mstar$ distribution of Class\,I appears flatter than the corresponding distribution for Class\,II objects.
\newline $-$ The $\mdisk \, vs. \, \mstar$ relation is flatter for young systems compared with the older ones. A large fraction of protostellar sources have the ratio $\mdisk/\mstar$ above 0.1, suggestive of propensity to gravitational instability.
    
Uniform samples of Class\,I and Class\,II protostars with identical initial conditions and the systematic analysis of more embedded sources are necessary to draw solid conclusions on the evolutionary path of YSOs and to be able to set the initial conditions for stars and planets formations. 
As shown in \citet[][]{fio21}, VLT/KMOS can be used on wider range of sources. Future observations by JWST will deliver information on photospheres with NIRSpec and eventually enable the investigation of the protostellar accretion rates for even more embedded sources with MIRI.

\begin{acknowledgments}
We thank the anonymous referee for their useful comments.
For this work we used ALMA archive data, we thus acknowledge 2016.1.01511.S, 2019.1.01813.S, and 2019.1.01792.S programs.
This work has made use of data from the European Space Agency (ESA) mission {\it Gaia} (\url{https://www.cosmos.esa.int/gaia}), processed by the {\it Gaia} Data Processing and Analysis Consortium (DPAC, \url{https://www.cosmos.esa.int/web/gaia/dpac/consortium}. 
Funding for the DPAC has been provided by national institutions, in particular the institutions participating in the {\it Gaia} Multilateral Agreement. 
This project has received funding from the European Research Council (ERC) under the European Union's Horizon 2020 Research \& Innovation Programme under grant agreement No 716155 (SACCRED) and by the European Union under the European Union’s Horizon Europe Research \& Innovation Programme 101039452 (WANDA) and 101039651 (DiscEvol). Views and opinions expressed are however those of the author(s) only and do not necessarily reflect those of the European Union or the European Research Council. Neither the European Union nor the granting authority can be held responsible for them.
GR acknowledges support from the Netherlands Organisation for Scientific Research (NWO, program number 016.Veni.192.233) and from an STFC Ernest Rutherford Fellowship (grant number ST/T003855/1).
\end{acknowledgments}

\bibliography{sample631}{}
\bibliographystyle{aasjournal}

\appendix

\begin{deluxetable*}{llcccccccc} 
\tablecaption{Stellar and accretion parameters of the Class\,II from CG10 computed in this work.\label{tab:classII}}
\tablewidth{0pt}
\tablehead{
\colhead{ID} & $T_{eff}$ & SpT & age & $A_ V$ & $\lacc$ & $\lstar$ &  $\mstar$ & $\rstar$ & $\log \macc$ \\
\colhead{  } &  K       &      & yr & mag    &$\lsun$ & $\lsun$ & $\msun$ & $\rsun$ & $\msun/$yr }
\startdata
08 &  5290 & G7 & $2.09 \times 10^{7}$ & $14.80 \pm 7.80$ & $0.01 \pm 0.01$ & $0.51 \pm 0.01$ & $1.11 \pm 0.01$ & $1.20 \pm 0.01$ & $-9.31 \pm 0.32$\\
12 & 5740 & G3 & $2.83 \times 10^6$& $1.47 \pm 1.00$& $0.77 \pm 0.13$& $2.78 \pm 0.13$& $1.48 \pm 0.14$ & $1.72 \pm 0.21$ & $-7.44 \pm 0.08$ \\
14 &  5290 & G7 & $ 2.09 \times 10^{7}$ & $ 2.43 \pm 0.07$ & $ 0.10 \pm 0.02$ & $0.88 \pm 0.01$ & $1.11 \pm 0.01$ & $1.20 \pm 0.01$ & $ -8.39 \pm 0.07$\\
35 &  9230 & A0 & $ 1.90\times 10^{7}$ & $ 4.64 \pm 0.36$ & $ 0.52 \pm 0.02$ & $19.18 \pm 0.02$ & $ 2.14\pm 0.01$ & $2.13 \pm 0.01$ & $ -7.68\pm 0.02$
\enddata
\end{deluxetable*}

\begin{deluxetable*}{llccccccc} 
\tablecaption{Stellar and accretion parameters of the FUors from \citet[][]{con10} computed in this work.\label{tab:fuors}}
\tablewidth{0pt}
\tablehead{
\colhead{ID} & age & $A_ V$ & $\lacc$ & $\lstar$ & $T_{eff}$ & $\mstar$ & $\rstar$ & $\macc$ \\
\colhead{  } &     & mag & $\lsun$ & $\lsun$ & K         &$\msun$ & $\rsun$ & $\msun/$yr  }
\startdata
04 & bl - 1Myr & $17 - 18$ & $ 0.03 -  0.04$ & $  0.28 - 0.30$ & $ 2818 - 2819$ & $0.15 - 0.13$ & $3.01 - 2.26$ & $-7.65 / -7.71$ \\
19 & bl - 1Myr & $14 - 11$ & $ 0.16 -  0.12$ & $  3.02 - 3.05$ & $ 3396 - 3719$ & $0.33 - 0.49$ & $4.77 - 3.64$ & $-7.07 / -7.48$ \\
32 & bl - 1Myr & $17 - 13$ & $168   -  109 $ & $125 -183$ & $ 4074 - 5754$ & $0.96 - 3.06$ & $7.02 - 5.69$ & $-4.32 / -5.11$ \\
37 & bl - 1Myr & $25 - 20$ & $25 - 15$ & $ 35 -46$ & $ 4074 - 5754$ & $0.96 - 3.06$ & $7.02 - 5.69$ & $-5.17 / -6.03$ \\
41 & bl - 1Myr & $30 - 23$ & $ 9.32 -  4.13$ & $ 56 -62$ & $ 4074 - 5754$ & $0.96 - 3.06$ & $7.02 - 5.69$ & $-5.65 / -6.61$ \\
43 & bl - 1Myr & $20 - 16$ & $ 0.08 -  0.05$ & $  9.11 - 9.15$ & $ 3852 - 4614$ & $0.70 - 1.77$ & $6.14 - 4.59$ & $-7.67 / -8.31$ 
\enddata
\end{deluxetable*}

\section{Class II}
\label{app:classII}
In this section we will present the accretion rates and stellar parameters for those source which were classified as Class\,II after  CG10 work, i.e. ID sources: 08, 12, 14, and 35.


These sources were analized by using the same routine we used for Class\,I but fixing the spectral type \citep[from][]{con10} and setting the age as a free parameter. Results are listed in Tab.\,\ref{tab:classII}, and are compatible with typical accretion rates and stellar parameters of CTTs.

\section{FUors}
\label{app:fuors}
We report in Tab.\,\ref{tab:fuors} the results we get for FUors sources, even if we are aware that our method is unlikely to work for such objects.

\section{Spectral Energy Distribution}
\label{app:sed}
\begin{figure*}
    \centering
    \includegraphics[width=0.3\textwidth]{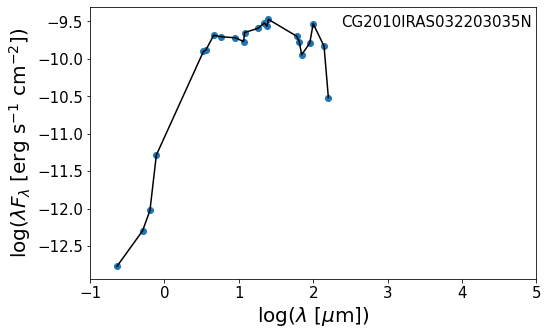}
    \includegraphics[width=0.3\textwidth]{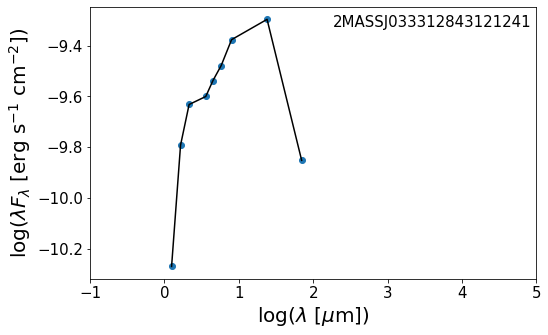}
    \includegraphics[width=0.3\textwidth]{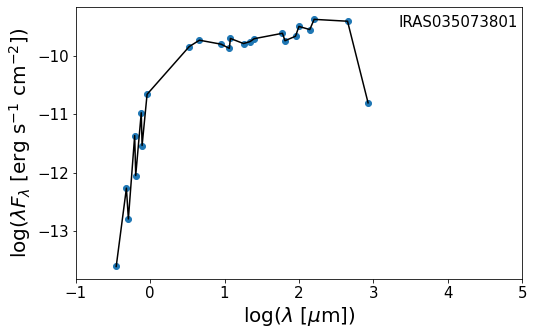}
    \includegraphics[width=0.3\textwidth]{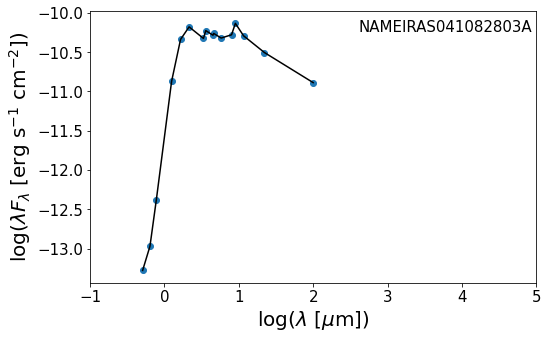}
    \includegraphics[width=0.3\textwidth]{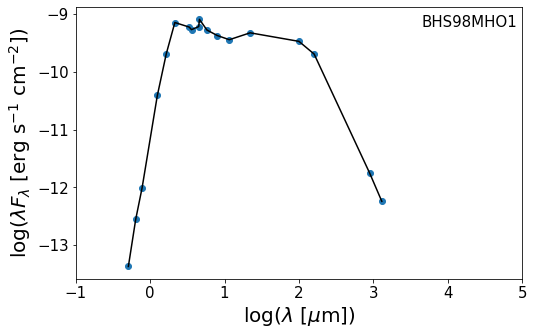}
    \includegraphics[width=0.3\textwidth]{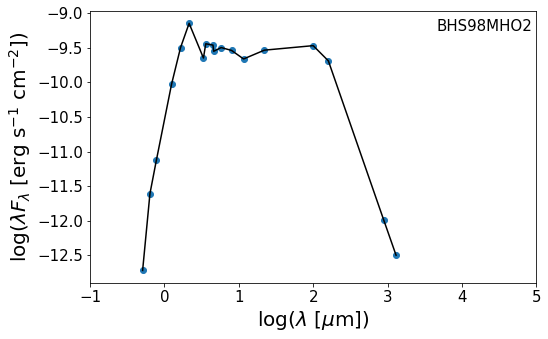}
    \includegraphics[width=0.3\textwidth]{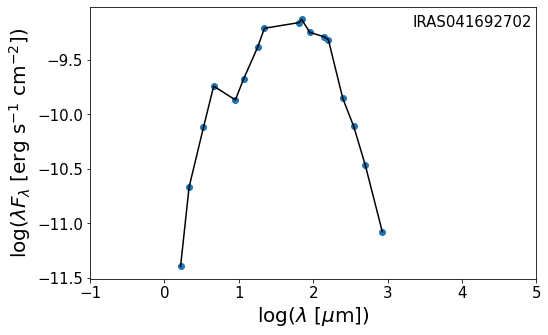}
    \includegraphics[width=0.3\textwidth]{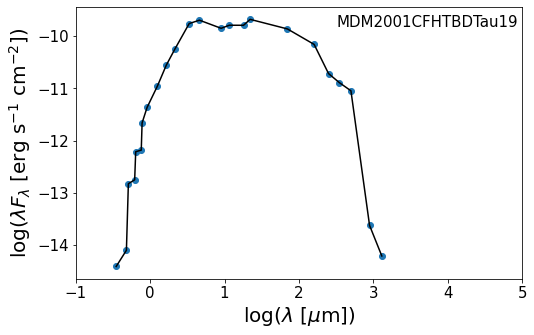}
    \includegraphics[width=0.3\textwidth]{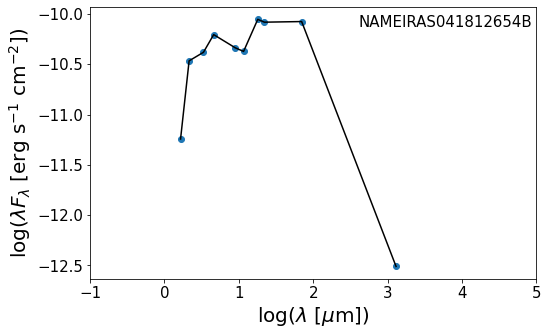}
    \includegraphics[width=0.3\textwidth]{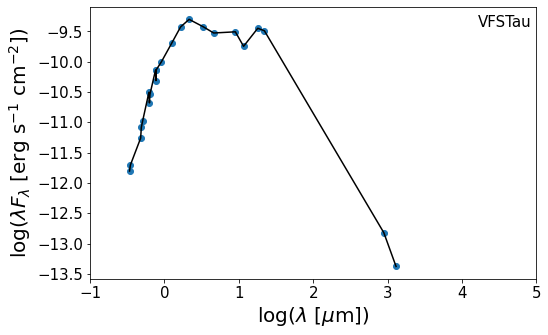}
    \includegraphics[width=0.3\textwidth]{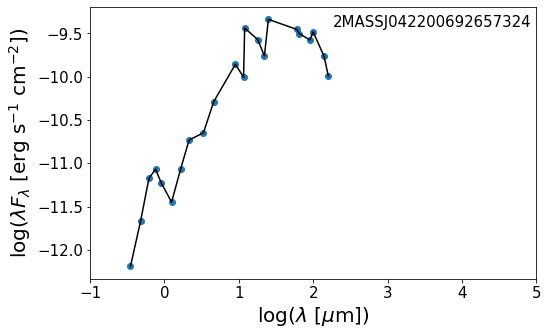}
    \includegraphics[width=0.3\textwidth]{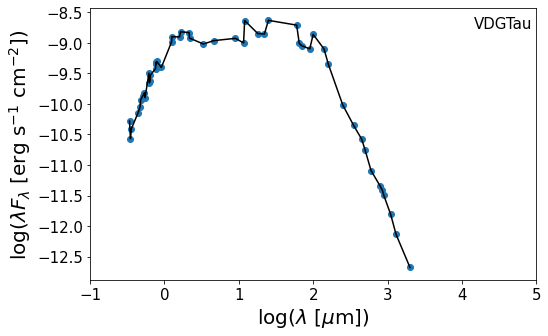}
    \includegraphics[width=0.3\textwidth]{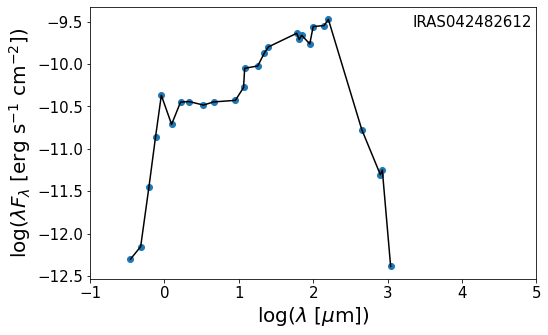}
    \includegraphics[width=0.3\textwidth]{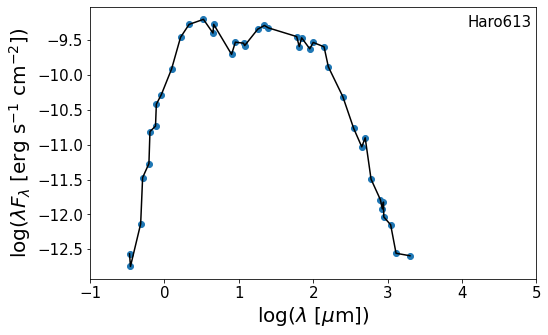}
    \includegraphics[width=0.3\textwidth]{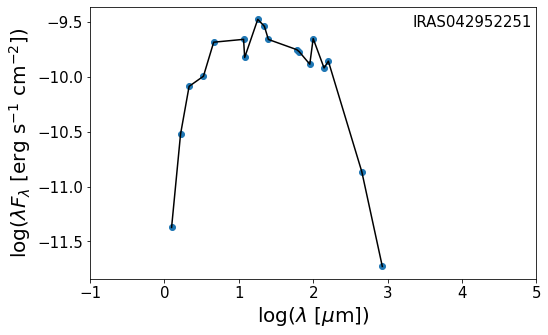}
    \includegraphics[width=0.3\textwidth]{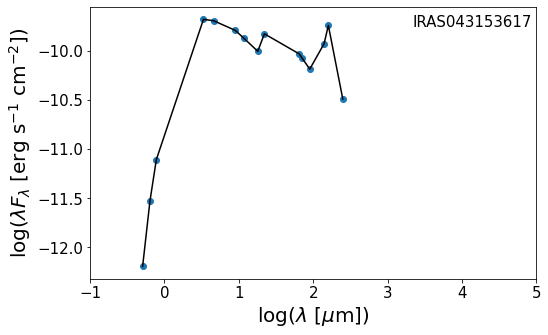}
    \includegraphics[width=0.3\textwidth]{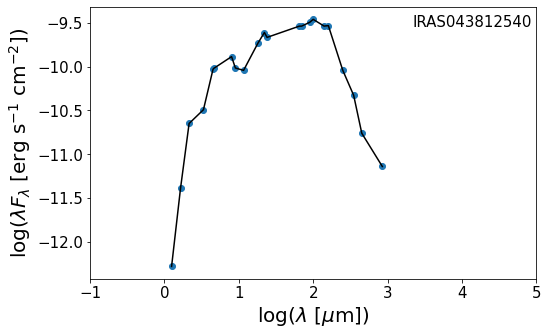}
    \includegraphics[width=0.3\textwidth]{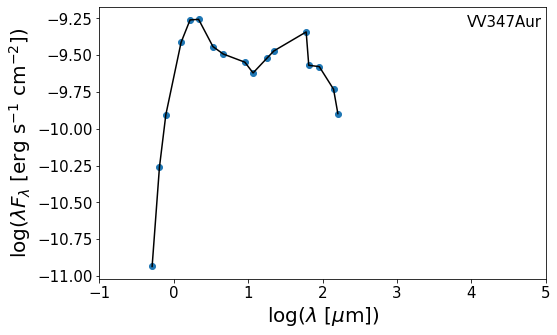}
    
    \caption{Spectral energy distributions of the YSOs.}
    \label{fig:Lbol1}
\end{figure*}

\begin{figure*}
    \centering
    \includegraphics[width=0.3\textwidth]{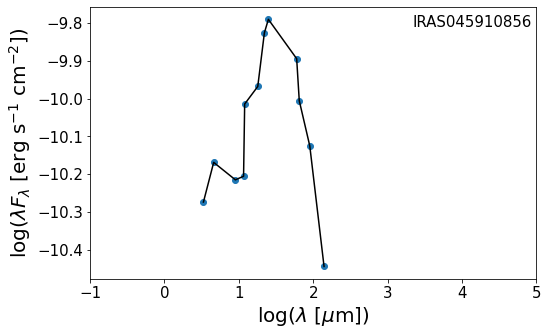}
    \includegraphics[width=0.3\textwidth]{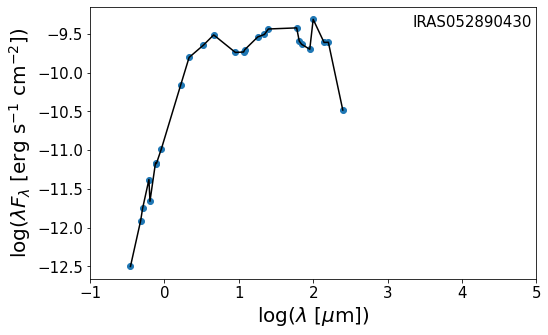}
    \includegraphics[width=0.3\textwidth]{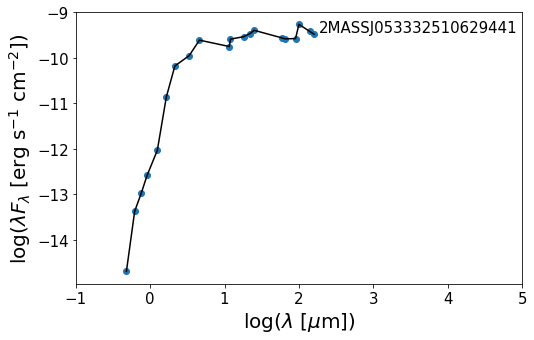}
    \includegraphics[width=0.3\textwidth]{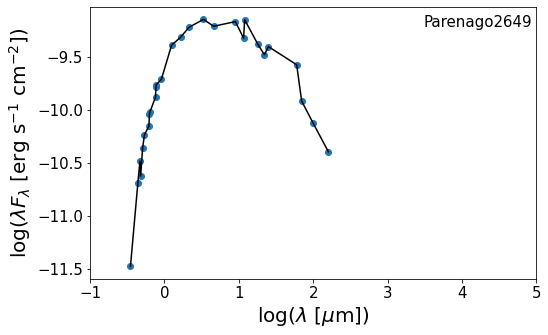}
    \includegraphics[width=0.3\textwidth]{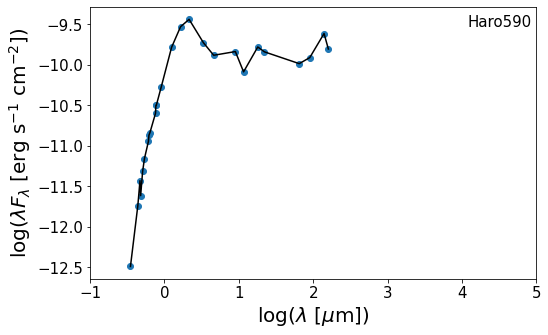}
    \includegraphics[width=0.3\textwidth]{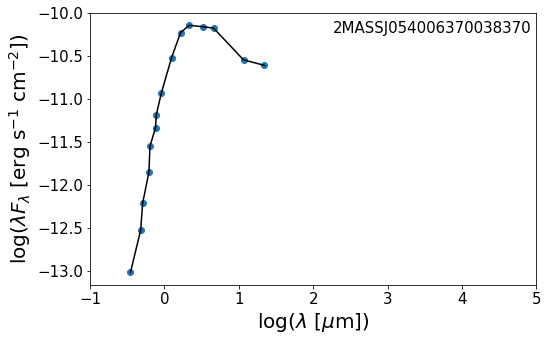}
    \includegraphics[width=0.3\textwidth]{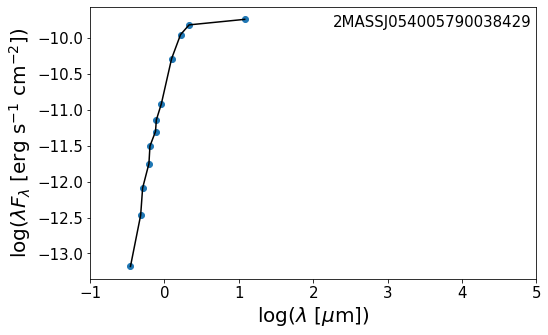}
    \includegraphics[width=0.3\textwidth]{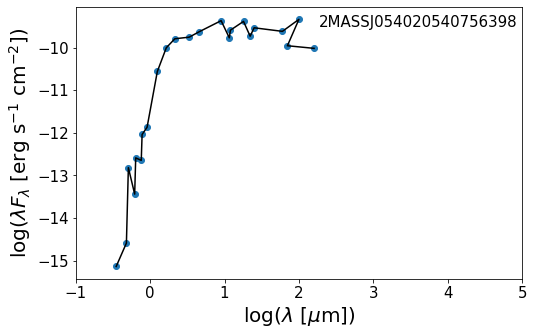}
    \includegraphics[width=0.3\textwidth]{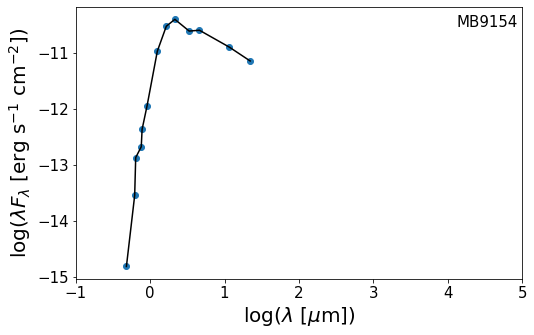}
    \includegraphics[width=0.3\textwidth]{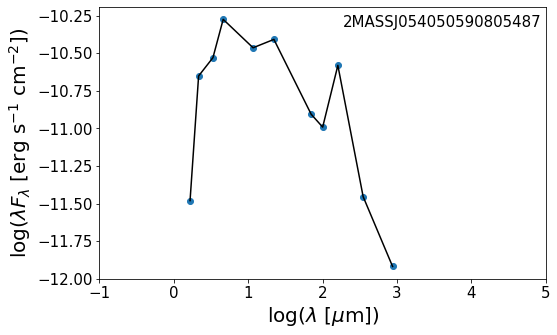}
    \includegraphics[width=0.3\textwidth]{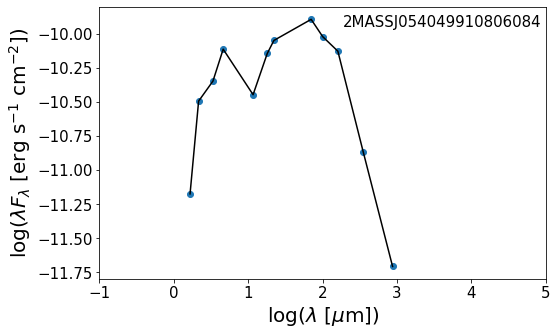}
    \includegraphics[width=0.3\textwidth]{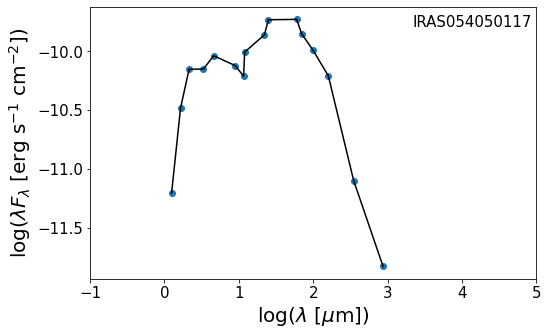}
    \includegraphics[width=0.3\textwidth]{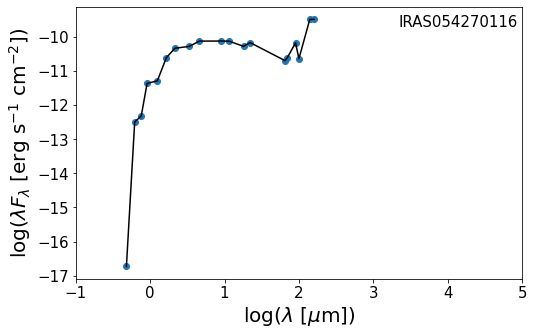}
    \includegraphics[width=0.3\textwidth]{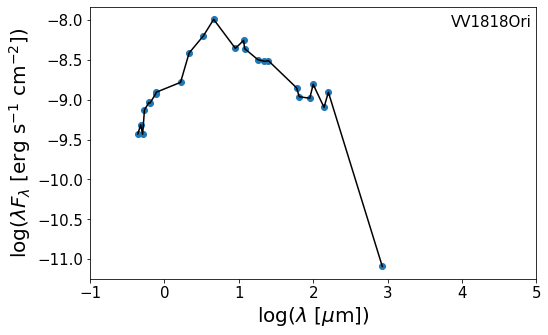}
    \includegraphics[width=0.3\textwidth]{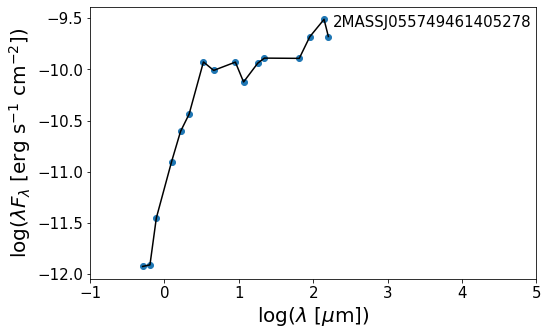}
    \includegraphics[width=0.3\textwidth]{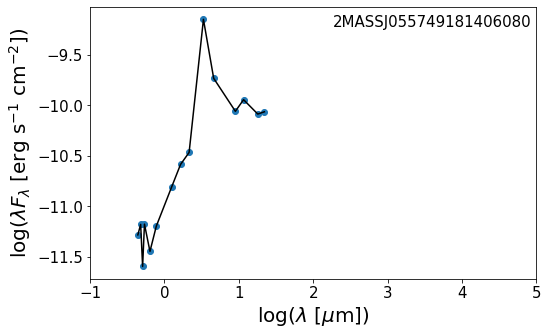}
    \includegraphics[width=0.3\textwidth]{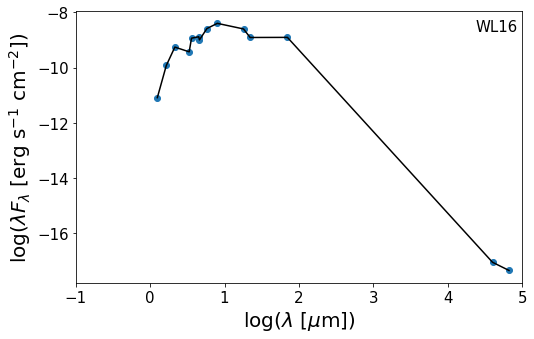}
    \includegraphics[width=0.3\textwidth]{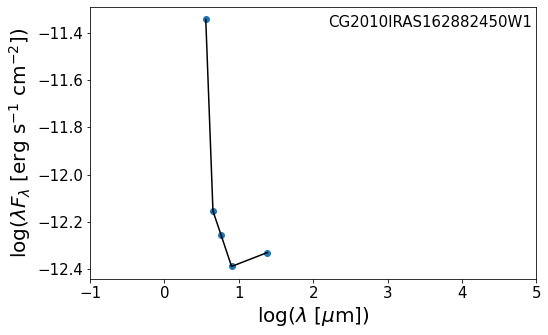}

    \caption{Spectral energy distributions of the YSOs.}
    \label{fig:Lbol2}
\end{figure*}

\begin{figure*}
    \centering
    \includegraphics[width=0.3\textwidth]{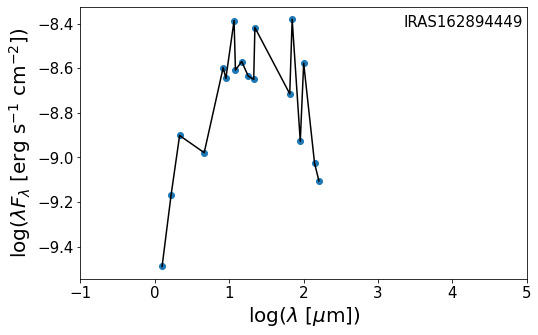}
    \includegraphics[width=0.3\textwidth]{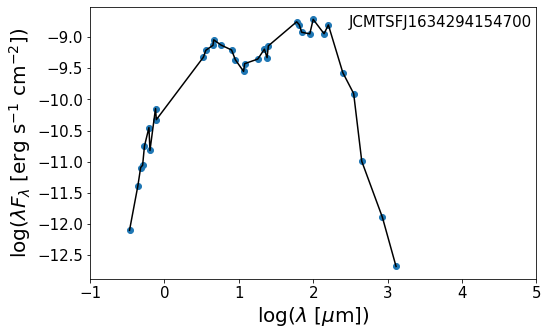}
    \includegraphics[width=0.3\textwidth]{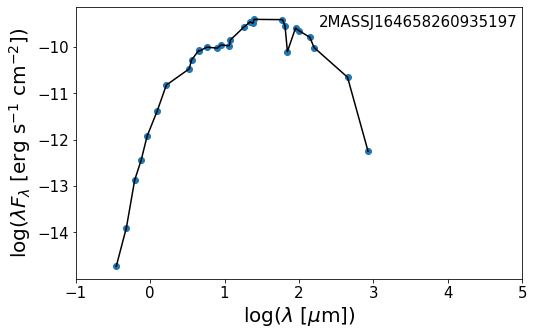}
    \includegraphics[width=0.3\textwidth]{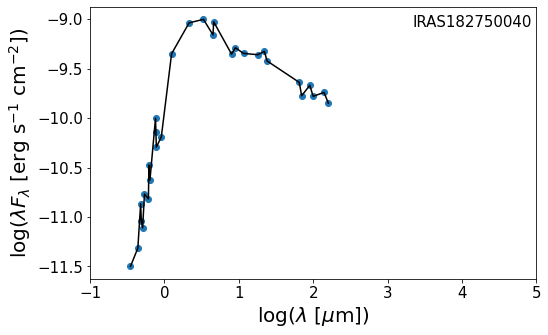}
    \includegraphics[width=0.3\textwidth]{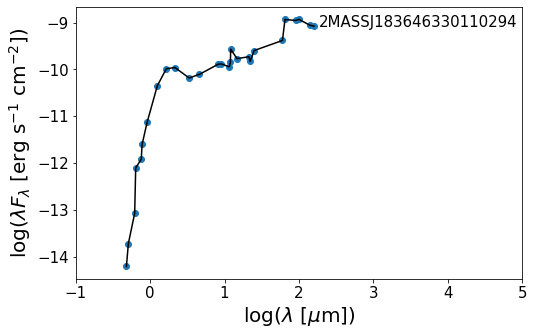}
    \includegraphics[width=0.3\textwidth]{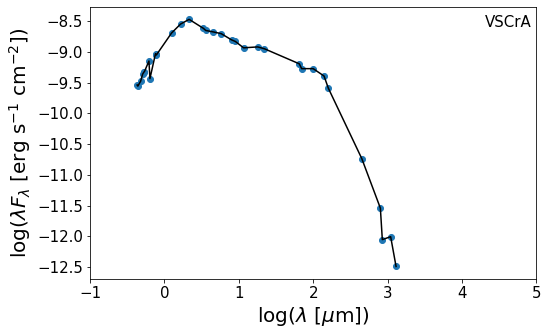}
    \includegraphics[width=0.3\textwidth]{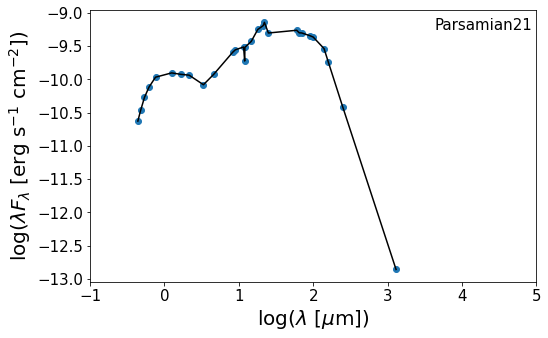}
    \includegraphics[width=0.3\textwidth]{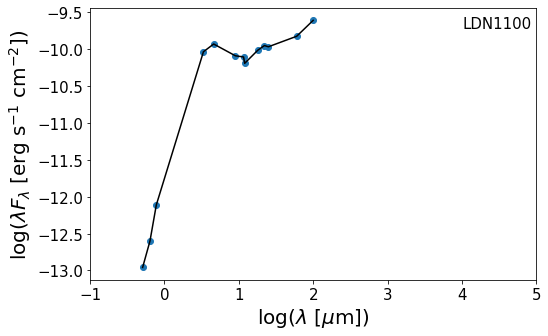}
    \includegraphics[width=0.3\textwidth]{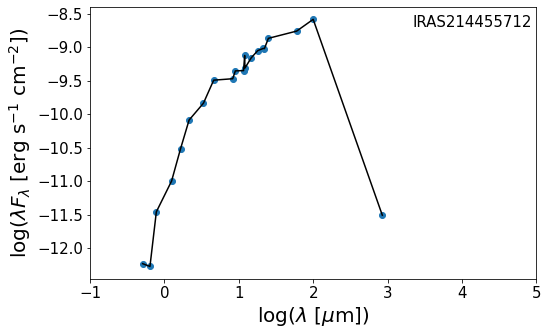}
    \includegraphics[width=0.3\textwidth]{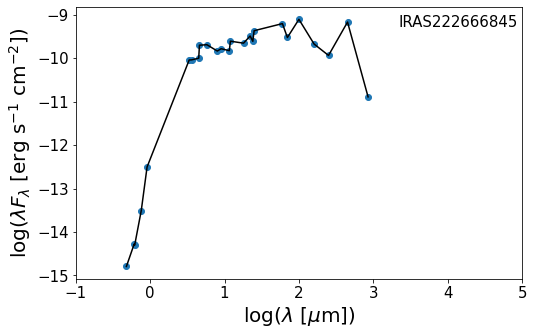}
    \includegraphics[width=0.3\textwidth]{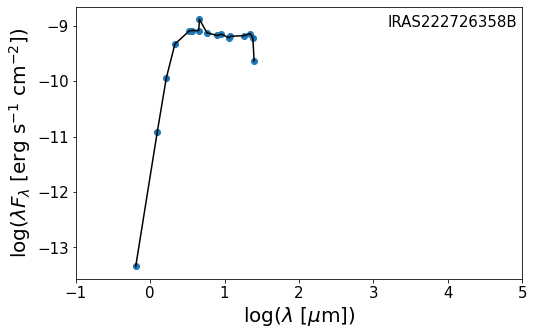}
    \includegraphics[width=0.3\textwidth]{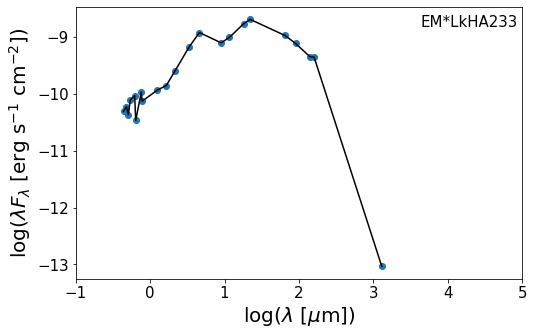}
    \includegraphics[width=0.3\textwidth]{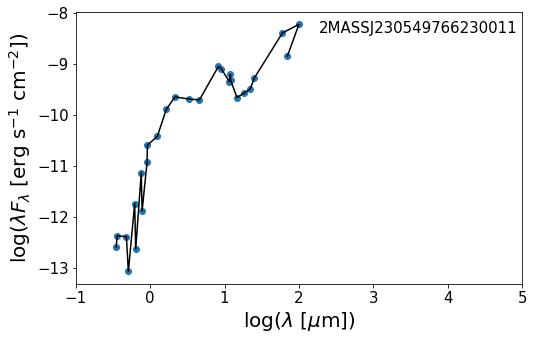}
    \includegraphics[width=0.3\textwidth]{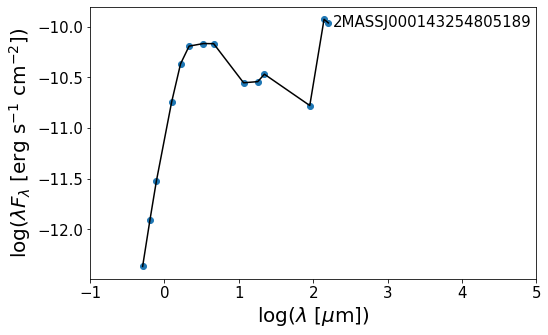}
    \caption{Spectral energy distributions of the YSOs.}
    \label{fig:Lbol3}
\end{figure*}

We present the Spectral Energy Distributions of our sample in Figs.\,\ref{fig:Lbol1}, \ref{fig:Lbol2}, and \ref{fig:Lbol3}.
We used the SEDBYS v.2.0 python-based package \citep{Davies2021} to automatically look for the photometry available in published catalogs; data from these specific resources were reviewed and included in the SEDs: \citet[][]{iras1988, hillenbrand1992, mannings1994, cutri2003, evans2003, andrews2007, difrancesco2008, kevork2009, ishihara2010, yamamura2010, bianchi2011, ahn2012, cutri2012, page2012, henden2015, shadab2015, ribas2017, gaia2018}.

In the following, we describe some interesting targets.

{\it 04108+2803A}. Suggested to be a reddened Class II source in \cite{Furlan.McClure.ea2008}. Separated by 21" from its B component, a much brighter in the IR 04108+2803B. Source A confirmed to be not detected longward of 100$\mu$m in {\it Herschel} maps.

{\it [MDM2001]CFHT-BD-Tau19}. Classified as brown dwarf. In \cite{Guieu.Dougados.ea2006} it presents very rich spectrum with prominent accretion lines, including optical lines, hinting at more evolved nature.

{\it DG Tau A}. Typically classified as T Tauri star \citep{Purser.Ainsworth.ea2018,Harrison.Looney.ea2019}. Index provided by \cite{con10} consistent with Class II.

{\it Haro 6-13}. Based on {\it Herschel} photometry classified as Class II \citep{Harrison.Looney.ea2019}

{\it 2MASSJ05405059-0805487, 2MASSJ05404991-0806084,IRAS05405-0117}. Based on {\it Herschel} photomerty classified as Flat spectrum sources \citep{Furlan.Fischer.ea2016}

{\it 2MASSJ05333251-0629441}. It is present in the SPIRE map but not listed in SPIRE Point Source Catalog. Shimajiri 2015 provides 1.1 mm flux.

{\it IRAS16289-4449}. It is present in the SPIRE map but not listed in SPIRE Point Source Catalog.

{\it Parenago 2649}. Non-detection in SPIRE map.

{\it WL 16}. Reclassified as Class II in \cite{Sadavoy.Stephens.ea2019} based on saturated IR spectrum, PAH detection and weak outflow.

\newpage
\startlongtable
\begin{deluxetable*}{llcccccc} 
\tablecaption{Disk masses for the Class I targets from gas kinematics.}
\label{tab:submm_fluxes}
\tablewidth{0pt}
\tablehead{
\colhead{ID} & Name &  Distance & $F_\lambda$  & $\lambda$ & $M_{\rm dust}$ & $M_{\rm disk}$ & ref\\
\colhead{}  & {}  &  pc & mJy  & mm & $\mearth$ & $\msun$ &  {}
}
\startdata
T1-01 		 & CG2010IRAS032203035N  & $219.8\pm16.2$      & $78.9 \pm 0.9 $	& 	1.3 & $ 167.5 \pm  24.8 $ & $0.0558 \pm 0.0083 $ & 1  \\
T1-02 		 & 2MASSJ033312843121241 & $319.5\pm23.7$     & $42.4 \pm 0.5 $ 	& 	1.3 & $ 190.2 \pm  28.3 $ & $0.0634 \pm 0.0094 $ &  2 \\
T1-05 		 & BHS98MHO1 			 & $134.0\pm7.0 $    & $217.0 \pm 0.8$ 	&	1.3 & $ 171.0 \pm  17.9 $ & $0.0570 \pm 0.0060 $ & 3 \\
T1-06 		 & BHS98MHO2 			 & $131.0\pm2.9$    & $133.0 \pm	0.8$ 	&	1.3 & $ 100.5 \pm  4.5 $ & $0.0335 \pm 0.0015 $ & 3 \\
T1-07 		 & IRAS041692702 		 & $129.5\pm 12.9$    & $200.0 \pm	50.0$ 	&	1.3 & $ 147.4 \pm  47.1 $ & $0.0491 \pm 0.0157 $ & 4 \\
T1-08        & MDM2001CFHTBDTau19    & $155.9 \pm 15.6 $  & $18.5 \pm	0.90$ 	&	0.9 & $ 7.2 \pm  1.5 $ & $0.0024 \pm 0.0005 $ &  5 \\
T1-10 		 & VFSTau 				 & $133.9\pm2.4$   & $2.3 \pm	0.14$ 	&	1.3 & $ 1.8 \pm  0.1 $ & $0.0006 \pm 0.0000 $ & 6\\
T1-11 		 & 2MASSJ042200692657324 & $133.9\pm2.4$   & $341.0 \pm	34.10$ 	&	0.9 & $ 97.3 \pm  10.3 $ & $0.0324 \pm 0.0034 $ & 7\\
T1-12$^{*}$  & VDGTau 				 & $125.3\pm1.9$   & $42.71 \pm	0.64$ 	&	3.0 & $ 324.6 \pm  11.0 $ & $0.1082 \pm 0.0037 $ & 8\\
T1-14$^{*}$  & Haro613 				 & $128.6\pm1.6$ & $25.0 \pm	0.08$ 	&	3.0 & $ 199.8 \pm  5.0 $ & $0.0666 \pm 0.0017 $ & 8 \\
T1-15 		 & IRAS042952251 		 & $160.76 \pm 16.1$ & $110.0 \pm	50.00$ 	&	1.3 & $ 124.9 \pm  62.0 $ & $0.0416 \pm 0.0207 $ &4\\
T1-17 		 & IRAS043812540 		 & $141.8 \pm 1.4$ & $27.9 \pm	0.20$ 	&	1.3 & $ 24.6 \pm 5.0 $ & $0.0082 \pm 0.0002 $ & 9\\
T1-22 		 & Parenago2649 		 & $398.5\pm2.5 $ & $15.0 \pm	3.00$ 	&	1.3 & $ 104.7 \pm  21.0 $ & $0.0349 \pm 0.0070 $ & 10\\
T1-28 		 & 2MASSJ054050590805487 & $440 \pm 44 $ & $23.5 \pm	0.72$ 	&	0.9 & $ 72.4 \pm  14.6 $ & $0.0241 \pm 0.0049 $ & 11\\
T1-29 		 & 2MASSJ054049910806084 & $440 \pm 44 $ & $3.3 \pm	0.56$ 	&	0.9 & $ 10.2 \pm  2.7 $ & $0.0034 \pm 0.0009 $ & 11\\
T1-30 		 & IRAS054050117 		 & $420 \pm 42 $ & $68.2 \pm	1.56$ 	&	0.9 & $ 191.5 \pm  38.6 $ & $0.0638 \pm 0.0129 $ & 11\\
T1-32 		 & VV1818Ori 			 & $633.4\pm22.7 $ & $3.3 \pm	0.07$ 	&	1.3 & $ 58.2 \pm  4.3 $ & $0.0194 \pm 0.0014 $ & 12\\
T1-35$^{*}$  & WL16 				& $138.4 \pm 2.6$ & $4.4 \pm	0.07$ 	&	1.3 & $ 3.7 \pm  0.2 $ & $0.0012 \pm 0.0001 $ & 13 \\
T1-37 		 & IRAS162894449 		& $350.3 \pm 2.5$	 & $27.0 \pm	3.00$ 	&	1.3 & $ 145.6 \pm  16.3 $ & $0.0485 \pm 0.0054 $ & 14 \\
T1-42 		 & VSCrA 				 & $160.5\pm1.8 $ & $271.0 \pm	0.80$ 	&	1.3 & $ 306.7 \pm  6.9 $ & $0.1022 \pm 0.0023 $ & 15 \\
T1-43 		 & Parsamian21 			 & $400 \pm 100	$ & $33.9 \pm	0.10$ 	&	1.3 & $ 214.4 \pm  107.2 $ & $0.0715 \pm 0.0357 $ & 14\\	
T2-09 & 03283968$+$3117321$^*$ 		 & $293 \pm 22$ 	   & $5.16 \pm 0.22$ 	& 	1.1 &	$12.8 \pm 0.8$	  & $ 0.0043 \pm 0.0003$    & 2 \\ 
T2-10 & J03285842$+$3122175$^*$ 	 & $293 \pm 22$ 	   & $4.57 \pm 0.16$ 	&	1.1 &	$11.3 \pm 0.6$    & $ 0.0038 \pm 0.0002$    & 2  \\ 
T2-11 & J03290149$+$3120208$^*$ 	 & $293 \pm 22$  	   & $3.10 \pm 0.64$    &   1.1 & 	$7.7  \pm 2.2$    & $ 0.0026 \pm 0.0007$ 	& 2 \\ 
T2-12 & SVS 13 (V512 Per) 		 & $293 \pm 22$ 	   & $257.1\pm 2.9$		&   1.3 &   $969.7\pm 15.5$   & $ 0.3232 \pm 0.00512$   & 1\\ 
T2-13 & LAL96 213 				 & $293 \pm 22$ 	   & $86.4 \pm 0.2$ 	&	1.3 & 	$318.5\pm 0.9 $   & $ 0.1063 \pm 0.0003 $   & 2\\ 
T2-17 & J03292003$+$3124076$^*$ 	 & $293 \pm 22$ 	   & $1.0  \pm 0.3$     &   1.3 &   $3.8  \pm 1.6 $   & $ 0.0013 \pm 0.0005 $   & 16\\ 
T2-01 & IRS2							 & 160 					& $616.9 \pm 1.0$		&   1.3 & $693.8\pm1.6$	 & $0.2313 \pm 0.0005$ & 15  \\
T2-02 &IRS5a 						 & 160 					& $522.0 \pm 1.3$	&   1.3 & $587.1\pm2.1$ 	 & $0.1957 \pm 0.0007$ & 15  \\
T2-04 & HH 100 IR						 & 160 					& $394.4 \pm 1.5$ 	&   1.3  & $443.5\pm8.0$ 	 &$0.1478 \pm 0.0027$ & 15\\
T2-06 & HH 26 IRS 					 & 450 					& $145.4 \pm 3.8 $ 	&   0.9	 &	$468.7\pm18.1 $  &$0.1562 \pm 0.0060$ & 11 \\
T2-07 & HH 34 IRS 					 & 460 					& $677.6 \pm 14.2$ 	&   0.9  &	$ 2282.7\pm67.7 $&$0.7609 \pm 0.0226$ & 11 \\
T2-08 & HH 46 IRS 					 & 450 					& $11.0 \pm 0.01$ 	&   3.0 &	$ 1126.8\pm14.5 $&$0.3756 \pm 0.0048$ & 17 \\
T3-01 & B335						 & $106 \pm 15$			& $28\pm0.2$		&   1.3 &   $13.8\pm3.9$ 		& $0.0041 \pm 0.0011$ & 18 \\ 
T3-02 & IRAS16253$-$2429				 & $144\pm9$ 			& $10.1\pm0.4$		&   1.3 &   $9.2\pm1.2$			& $0.0028\pm0.0004$ & 19 \\
T3-03 & VLA1623A					 & $144\pm9$ 			& $141.8 \pm 5.9$	&   1.3 &   $129 \pm 16$		& $0.0388\pm0.0049$ & 20 \\
T3-04 & IRAS15389$-$3559				 & $155\pm4$			& $6.98\pm0.12$	   &   1.2 &   $5.9\pm 0.3$		& $0.0018\pm0.0001$ & 21 \\
T3-05 & Lupus3$-$MMS 				 & $162\pm3$			& $185.1\pm0.2$	&      1.3 &   $213\pm 8$			& $0.0634\pm0.0024$ & 22 \\
T3-06 & L1455$-$IRS1 				 & $293\pm22$			& $46.5\pm0.5$	&      1.3 &   $175\pm 26$			& $0.0526\pm0.0048$ & 1 \\ 
T3-07 & IRAS4A2   					 & $293\pm22$			& $434.2\pm7.9$	   &   1.3 &   $1637\pm148$		& $0.4916\pm0.0444$ & 1 \\ 
T3-08 & L1157   					 & $352\pm19$			& $181\pm30$	   &   1.3 &   $985\pm195$			& $0.2956\pm0.0586$ & 23 \\ 
T3-09 & HH 212   					 & $400\pm40$			& $186\pm4$	  	   &   0.9 	&   $474\pm95$			& $0.1423\pm0.0285$ & 11 \\ 
T3-10 & L1527   					 & $141\pm9$			& $23\pm0.11$	   &   3.0 	&   $241\pm39$			& $0.0723\pm0.0117$ & 24 \\ 
T3-11 & L1551$-$IRS5    				 & $141\pm9$			& $602.2\pm55.9$	   &  1.3 	&   $526\pm83$			& $0.1580\pm0.0025$ & 25 \\ 
T3-12 & TMC1A    				 	 & $141\pm9$			& $240 \pm 24$	   			&  1.3 	&   $210 \pm 34$			& $0.0631\pm0.0102$ & 26 \\ 
T3-13 & L1489 IRS    				 & $141\pm9$			& $59\pm5$	   			&  1.3 	&   $52 \pm 8$			& $0.0155\pm0.0023$ & 27 \\ 
T3-13 & L1551$-$NE    				 & $141\pm9$			& $600\pm1$	   		&  0.9 	&   $187\pm24$			& $0.0562\pm0.0072$  & 28 \\ 
T3-14 & WL12						 & $144\pm9$			& $71.1 \pm 0.4$    & 1.3 	&	$62.1 \pm 7.9$ 		&$0.0187\pm	0.0024$ & 15 \\
T3-15 & Elias29						 & $144\pm9$			& $17.2 \pm 0.2$    &  1.3 	&	$15.6 \pm 2.0$ 		&$0.0047\pm0.0006$ & 13 \\
T3-16 & IRS63 						 & $144\pm9$			& $335.\pm18$    &  1.3 	&	$305\pm38$ 		&$0.0916\pm	0.0114$ & 29 \\
T3-17 & IRS43 						 & $144\pm9$			& $16.9\pm0.3$		&  1.3 		& $15.4\pm1.9$		&	$0.0046\pm0.0006$ & 13 \\
T3-18 & RCrA IRS7B  				 & $155 \pm 4$			& $372\pm1$		&  1.3 		& $393\pm17$		&	$0.1180\pm0.0051$ & 15 \\
T3-19 & HH 111 						 & $411 \pm 41$			& $285\pm40$	&  1.3  &$187\pm24$ &$0.0562 \pm 0.0072$ & 30\\
T3-20 & EC53 						 & $436\pm9$			& $139.7\pm1.2$	&  0.9  &$381\pm16$ &$0.1144\pm0.0048$ & 31 \\
\enddata

\tablecomments{1 - \cite{Tobin.Looney.ea2018}, 2 - \cite{tyc20}, 3 - \cite{Akeson.Jensen2014},4 - \cite[SMA;][]{Sheehan.Eisner2017}, 5 - ALMA: 2016.1.01511.S, 6 -\cite{Akeson.Jensen.ea2019}, 7 -\cite{Villenave.Menard.ea2020}, 8 -\cite{Harrison.Looney.ea2019}, 9 -\cite{vantHoff.Harsono.ea2020}, 10 - ALMA: 2019.1.01813.S, 11 -\cite{tob20}, 12 -\cite{Dutta.Lee.ea2020}, 13 -\cite{Sadavoy.Stephens.ea2019}, 14 -\cite{Kospal.CruzSaenzdeMiera.ea2021} 15 - ALMA: 2019.1.01792.S, 16 - \cite{Yang.Sakai.ea2021}, 17 -\cite{Zhang.Arce.ea2016}, 18 - \cite{Bjerkeli.Ramsey.ea2019}, 19 -\cite{Hsieh.Hirano.ea2019}, 20 - \cite{Sadavoy.Myers.ea2018}, 21 -\cite{Okoda.Oya.ea2018}, 
 22 -\cite{Yen.Koch.ea2017},23 -\cite{Chiang.Looney.ea2012},24 -\cite{Nakatani.Liu.ea2020},25 -\cite[ circumbinary disk, flux averaged of natural and uniform weighting][]{CruzSaenzdeMiera.Kospal.ea2019},26-\cite{Harsono.vanderWiel.ea2021}, 27-\cite{Sai.Ohashi.ea2020}, 28- \cite[flux of circumbinary disk][]{Takakuwa.Saigo.ea2017a},29-\cite{SeguraCox.Schmiedeke.ea2020},30-\cite{Lee2010}, 31-\cite{Lee.Lee.ea2020}.}
\end{deluxetable*}

\end{document}